\documentclass[11pt]{article}

\usepackage{amsfonts}
\usepackage{eucal}
\usepackage{amssymb,epsfig,latexsym}
\usepackage{graphicx,amsmath,amssymb,latexsym,psfrag,url}
\usepackage{multirow}
\usepackage{caption}

\usepackage{subcaption}
\usepackage{xspace,epsfig}
\usepackage{graphics,graphicx}

\newcommand{\R}{\mathbb{R}}

\newcommand{\C}{\mathbb{C}}
\newcommand{\EE}{\mathbb{E}}
\newcommand{\PP}{\mathbb{P}}

\newtheorem{Theorem}{Theorem}[section]
\newtheorem{Proposition}[Theorem]{Proposition}

\newtheorem{Corollary}[Theorem]{Corollary}
\newtheorem{Lemma}[Theorem]{Lemma}

\def\ind{{\rm 1\hspace{-0.90ex}1}}

\begin{document}
\title{Large deviations of the interference in the Ginibre network model}
\author{Giovanni Luca
Torrisi\thanks{Istituto per le Applicazioni del Calcolo "Mauro
Picone", CNR, Via dei Taurini 19, I-00185 Roma, Italia. e-mail:
 \tt{torrisi@iac.rm.cnr.it}}\,\, and\,\,
Emilio
Leonardi\thanks{Dipartimento di Elettronica, Politecnico di
Torino, Corso Duca degli Abruzzi 24, I-10129 Torino, Italia.
e-mail: \tt{leonardi@polito.it}}
}

\date{}
\maketitle

\noindent\emph{To appear  in  INFORMS-APS Stochastic Systems}\\
 \url{http://www.i-journals.org/ssy/}

\begin{abstract}
Under different assumptions on the distribution of the fading random variables, we derive large deviation estimates for the tail of the interference in a wireless network model whose nodes are placed, over a bounded region of the plane, according to the $\beta$-Ginibre process, $0<\beta\leq 1$. The family of $\beta$-Ginibre processes is formed by determinantal point processes, with different degree of repulsiveness, which converge in law to a homogeneous Poisson process, as $\beta \to 0$.
In this sense the Poisson network model may be considered as the limiting uncorrelated case of the $\beta$-Ginibre network model. Our results indicate the existence of two different regimes. When the fading random variables are bounded or Weibull superexponential, large values of the interference are typically originated
by the sum of several equivalent interfering contributions due to nodes in the vicinity of the receiver.

In this case, the tail of the interference has, on the log-scale, the same asymptotic behavior  for any value of $0<\beta\le 1$, but it differs  from the asymptotic behavior of the tail of the interference in the Poisson network model (again on a log-scale)~\cite{ganesh}.

When the fading random variables are exponential or subexponential, instead, large values of the interference are typically originated by a single dominating interferer node and, on the log-scale, the asymptotic behavior of the  tail of the interference is essentially insensitive to the distribution of the nodes. As a consequence, on the log-scale, the asymptotic behavior of the tail of the interference in any $\beta$-Ginibre network model, $0<\beta\le 1$, is the same as in the Poisson network model.
\end{abstract}





\section{Introduction}\label{sec:int}

An important performance index in a wireless network is the so-called outage (or success) probability,
which measures the reliability degree  of communications channels  established  between each  transmitter and its associated receiver.
The outage probability is mainly determined by the
mutual interference among simultaneous transmissions over the same physical channel~\cite{Gupta-Kumar,HAW08,bibbiatse,WAJ07,WYAV05}. In
the last years a huge effort has been  devoted
to characterize the interference produced by transmitting nodes
operating over the same
channel~\cite{libro_baccelli1,libro_baccelli2,BBM06,DBT05,DFMMT06,ganesh, GH09,GAH,
Baccelli:expansion,GGH,HG09,DeVeciana:subm,Baccelli_CSMA,privtor,leonardi}.
Most of these works, however, focused on networks in which
transmitting nodes are either distributed according to a
homogeneous Poisson process or, in a few cases, located on a
perfectly regular grid.

Although the Poisson assumption offers many analytical
advantages, it appears rather unrealistic in many cases, since it
neglects the correlations among the positions of different
transmitters, possibly resulting from the application of smart
scheduling policies or intelligent network planning techniques.
The assumption that transmitting nodes are located on a perfectly
regular grid is unrealistic too, since it does not capture the
effects of environmental constraints that prevent network planners
from placing wireless access points regularly spaced.

In many  practical situations, the set of nodes that transmit
simultaneously over the same channel may be thought as a point
process of repulsive nature, i.e. a point process whose points are negatively
correlated.
However, only very recently, the research community has started
 investigating  the mathematical properties of wireless network models in which transmitting nodes
are distributed according to general point
processes~\cite{nostro_CSMA,GAH,GH09,Baccelli:expansion,GGH,DeVeciana:subm,Baccelli_CSMA,privtor}.

Under various assumptions on the distribution of the fading random variables (i.e. signal powers)
and on the attenuation function, a first attempt to analyze the performance of a network in which
nodes locations are modeled as a general stationary and isotropic point process
has been carried out in \cite{GAH}, \cite{Baccelli:expansion} and \cite{GGH}.
In \cite{GAH} and \cite{GGH} the authors study the asymptotic behavior of the outage
probability as the intensity of the nodes goes to zero.
In \cite{Baccelli:expansion}, instead, the outage probability of the
network is approximated using the factorial moment expansion of
functionals of point processes and the proposed moment expansion can be
successfully applied when the joint intensities of the underlying point process
can be efficiently computed.
In \cite{nostro_CSMA,DeVeciana:subm,Baccelli_CSMA}, the authors propose different methodologies to estimate the outage probability of networks
in which the nodes are distributed according to a
Mat\'{e}rn hard-core  process. At last, in~\cite{GH09}
 authors characterize the outage probability of wireless networks in which
 nodes  are  distributed according
to attractive Poisson cluster processes, such as Neyman-Scott, Thomas and Mat\'ern
point processes and fading variables are exponentially distributed.

This paper may be considered as a natural extension of the study started
in \cite{ganesh}, where large deviation estimates for the
interference in the Poisson network model have been provided, under various assumptions on the
distribution of the fading random variables. Here we move a step forward
targeting networks in which the nodes are placed according to repulsive
point processes. Our main findings can be summarized as follows.
When the fading random variables are bounded or Weibull superexponential and the nodes are placed according to
the $\beta$-Ginibre process, $0<\beta\leq 1$, we derive the large deviations of the interference
by relating the tail of the interference with the number of points
falling in the proximity of the receiver. Our results show that, on the log-scale, the tail of the interference
exhibits the same asymptotic behavior for any value of $\beta\in (0,1]$.
At the same time, our results indicate that, on the log-scale,
the asymptotic behavior of the tail of the interference in the $\beta$-Ginibre network model,
$0<\beta\leq 1$, and the asymptotic behavior of the tail of the interference
in the Poisson network model are different. Since the Poisson process is the weak limit of the
$\beta$-Ginibre process, as $\beta \to 0$,
this enlightens a discontinuous behavior of the tail of the interference with respect to the
convergence in law. When the fading random variables are exponential or subexponential,
we prove that, on the log-scale, the asymptotic behavior of the tail of the interference
is insensitive to the distribution of the nodes,
as long as the number of nodes
is guaranteed to be light-tailed. Such insensitivity property
descends from the fact that
large values of the interference
are typically originated by a single dominant interferer node.

From a mathematical point of view,
the analysis of the $\beta$-Ginibre network model, $0<\beta\leq 1$, carried out in this paper differs from
the analysis of the Poisson network model studied in \cite{ganesh}, since we can not anymore resort on the independence
properties of the Poisson process. This difficulty is circumvented by combining ad hoc arguments, that leverage the
specific structure of the $\beta$-Ginibre process, $0<\beta\leq 1$, and the properties of subexponential distributions.

The paper is organized as follows. In Section \ref{sec:prel} we
describe the system model. In Section \ref{sec:prel0} we give some preliminaries on large deviations,
determinantal processes and $\beta$-Ginibre processes, $0<\beta\leq 1$.
The statistical assumptions on the model are provided in Section \ref{sec:ginibre}. In Sections
\ref{sec:ldptrunbsfad} and \ref{sec:ldptruncWeibfad} we derive
the large deviations of the interference in the $\beta$-Ginibre network model, $0<\beta\leq 1$, when
the fading random variables are bounded and Weibull superexponential, respectively. In Section
\ref{sec:ldpexp} we provide the large deviations
of the interference in more general network models when the signal powers are
exponential or subexponential. In Section \ref{sec:conc} we summarize the main
findings of this paper.
We include an Appendix where some technical results are proved.

\section{The system model}\label{sec:prel}

We consider the following simple model of wireless
network, which accounts for interference among different simultaneous transmissions.
Transmitting nodes (antennas) are distributed according to a simple (i.e. without multiple
points) point process $\mathbf{N}\equiv\{Y_i\}_{i\geq 1}$ on the plane.
One of the points of $\mathbf{N}$ is placed at
the origin, say $O$. A tagged receiver is then added at $y\in \R^2$.

We suppose that the useful signal emitted by the node at the origin is received at $y$ with power
 $Z_0 L(y)$, where $L:\R^2\to (0,\infty)$ is a  non increasing function called
attenuation function, and $Z_0$ is a random term modeling the effects of the fading.
Similarly, we assume that the interfering signal emitted by the node at $Y_i\neq O$ is received at $y$ with power $Z_iL(y-Y_i)$.
We suppose that the fading random variables $Z_i$
are non-negative, independent and identically distributed and independent of
$\{Y_i\}_{i\geq 1}$. Finally, we denote by $w>0$ the average thermal power noise at the receiver.

Let $\{X_i\}_{i\geq 1}$ denote the points of the point process
$\mathbf{N}\setminus\{O\}\,|\,O\in\mathbf N$ (the law of this process
is the so-called reduced Palm probability of $\mathbf N$ at the origin, see
e.g. \cite{daley1}.) We shall analyze the interference due to simultaneous transmissions
of nodes falling in a measurable and bounded region $\Lambda$ of the plane that contains  both $O$ and $y$ in its  interior. Assuming that all the random
quantities considered above are defined on the same probability space $(\Omega,\mathcal{F},\PP)$,
we define the interference by
\[
I_{\Lambda}=\sum_{i\geq 1}Z_iL(y-X_i)\ind_{\Lambda}(X_i)
\]
where, with a slight abuse of notation, we have still
denoted by $Z_i$ the fading random variable associated to the transmission of the node at $X_i$.
Here the symbol $\ind_{\Lambda}$
denotes the indicator function of the set $\Lambda$.

The tail of the interference is tightly related to
the probability of successfully decoding the signal from
the transmitter at the origin. Indeed, depending on  the adopted modulation and encoding scheme,
the receiver at $y$  can successfully
decode the signal from the transmitter at $O$ if the Signal to Interference plus Noise
Ratio (SINR) at the receiver is greater than a given threshold, say $\tau>0$ (which depends on the adopted scheme.)
In other words,  the success probability is given by
\begin{align}
\PP(\mathrm{SINR}>\tau)\qquad\text{where}\qquad\mathrm{SINR}=\frac{Z_0 L(y)}{w+I_\Lambda}.\nonumber
\end{align}

The relationship between  the tail of $I_\Lambda$ and the success probability
is highlighted by the following relation
\begin{equation*}
 \PP(\mathrm{SINR}>\tau \mid Z_0=z) =\PP\left(I_\Lambda<\frac{z L(y)}{\tau}-w\right).
\end{equation*}

\section{Preliminaries}\label{sec:prel0}

In this section, first we recall the notion of large deviation
principle and subexponential distribution (the reader is directed
to \cite{dembo} for an introduction to large deviations theory and
to \cite{asmussen} for more insight into heavy-tailed random
variables), second we recall the definition of determinantal
process, explain its repulsive nature and provide the definition
of $\beta$-Ginibre process, $0<\beta\leq 1$ (the reader is referred
to \cite{daley}, \cite{daley1} and \cite{moller} for notions of
point processes theory, to \cite{hough} for more insight into
determinantal processes and to \cite{blank} and \cite{brezis} for
notions of functional analysis.)

\subsection{Large deviation principles}\label{sec:ldp}

A family of probability measures
$\{\mu_\varepsilon\}_{\varepsilon>0}$ on
$([0,\infty),\mathcal{B}([0,\infty)))$ obeys a large deviation
principle (LDP) with rate function $I$ and speed $v$ if
$I:[0,\infty)\rightarrow [0,\infty]$ is a lower semi-continuous
function, $v:(0,\infty)\rightarrow (0,\infty)$ is a measurable
function which diverges to infinity at the origin, and the
following inequalities hold for every Borel set
$B\in\mathcal{B}([0,\infty))$:
\[
-\inf_{x\in B^\circ }I(x)\leq\liminf _{\varepsilon\rightarrow
0}\frac{1}{v(\varepsilon)}\log\mu_\varepsilon (B)\leq\limsup
_{\varepsilon\rightarrow
0}\frac{1}{v(\varepsilon)}\log\mu_{\varepsilon} (B)\leq -\inf
_{x\in \overline{B}}I(x),
\]
where $B^\circ$ denotes the interior of $B$ and $\overline{B}$
denotes the closure of $B$. Similarly, we say that a family of
$[0,\infty)$-valued random variables
$\{V_\varepsilon\}_{\varepsilon>0}$ obeys an LDP if
$\{\mu_\varepsilon\}_{\varepsilon>0}$ obeys an LDP and
$\mu_\varepsilon (\cdot)=P(V_\varepsilon\in\cdot)$. We point out
that the lower semi-continuity of $I$ means that its level sets:
\[
\{x\in [0,\infty): I(x)\leq a\},\quad\text{$a\geq 0$,}
\]
are closed; when the level sets are compact the rate function $I$
is said to be good.

In this paper we shall use the following criterion to provide the
large deviations of a non-negative family of random variables. Although its
proof is quite standard, we give it in the Appendix for the sake of completeness.
\begin{Proposition}\label{pro:crit}
Let $I:[0,\infty)\to [0,\infty)$ be an increasing function which
is continuous on $(0,\infty)$ and such that $I(0)=0$ and let
$v:(0,\infty)\rightarrow (0,\infty)$ be a measurable function
which diverges to infinity at the origin. If
$\{V_\varepsilon\}_{\varepsilon>0}$ is a family of non-negative
random variables such that $V_\varepsilon\downarrow 0$ and, for
any $x\geq 0$,
\[
\limsup_{\varepsilon\to
0}\frac{1}{v(\varepsilon)}\log\PP(V_\varepsilon\geq x)\leq -I(x)
\]
and
\[
\liminf_{\varepsilon\to
0}\frac{1}{v(\varepsilon)}\log\PP(V_\varepsilon>x)\geq -I(x),
\]
then the family of random variables
$\{V_\varepsilon\}_{\varepsilon>0}$ obeys an LDP on $[0,\infty)$
with speed $v$ and rate function $I$.
\end{Proposition}

A random variable $Z$ is called subexponential if it has support
on $(0,\infty)$ and
\[
\lim_{x\to\infty}\frac{\overline{F^{*2}}(x)}{\overline{F}(x)}=2,
\]
where $F(x)=\PP(Z\leq x)$, $\overline{F}(x)=\PP(Z>x)$ and $F^{*2}$
is the two-fold convolution of $F$.

Finally, we fix some notation. Let $f$ and $g$ be two real-valued
functions defined on some subset of $\R$. We write $f(x)=O(g(x))$
if there exist constants $M>0$ and $x_0\in\R$ such that
$|f(x)|\leq M|g(x)|$ for all $x>x_0$. We write $f(x)=o(g(x))$ if
for any $\varepsilon>0$ there exists $x_0\in\R$ such that
$|f(x)|\leq\varepsilon|g(x)|$ for all $x>x_0$. We write $f(x)\sim
g(x)$ if $\lim_{x\to\infty}f(x)/g(x)=1$. For any complex number
$z\in\C$, we denote by $\overline z$ its complex conjugate. For
any $x_0\in\mathbb{R}^2$ or $\mathbb C$, we denote by $b(x_0,r)$
the closed ball in $\R^2$ or $\mathbb C$ of radius $r>0$
centered at $x_0$. For any $x\geq 0$, we denote by $[x]$ the
biggest integer not exceeding $x$.

\subsection{Determinantal processes and their repulsive nature}\label{sec:determinantal}

We start recalling the notion of joint intensities (or $k$th order
product density functions) of a point process on the complex field.
Let $S\subseteq\C$ be a measurable set, $\lambda$ a
Radon measure on $S$ and $\mathbf{N}\equiv\{Y_i\}_{i\geq 1}$ a
simple point process on $S$. The joint intensities of $\mathbf N$
with respect to $\lambda$ are measurable functions (if any exist)
$\rho^{(k)}:S^k\to[0,\infty)$, $k\geq 1$, such that for any family
of mutually disjoint subsets $\Lambda_1,\ldots,\Lambda_k$ of $S$
\[
\EE\left[\prod_{j=1}^{k}\left(\sum_{i\geq
1}\ind_{\Lambda_j}(Y_i)\right)\right]=\int_{\prod_{j=1}^k
\Lambda_j}\rho^{(k)}(x_1,\ldots,x_k)\,\lambda(\mathrm{d}x_1)\ldots\lambda(\mathrm{d}x_k).
\]
In addition, we require that $\rho^{(k)}(x_1,\ldots,x_k)$ vanishes
if $x_h=x_k$ for some $h\neq k$. Intuitively, for any
pairwise distinct points $x_1,\ldots,x_k\in S$,
$\rho^{(k)}(x_1,\ldots,x_k)\,\lambda(\mathrm{d}x_1)\ldots\lambda(\mathrm{d}x_k)$
is the probability that, for each $i=1,\ldots,k$, $\mathbf N$ has a
point in an infinitesimally small region around $x_i$ of volume
$\lambda(\mathrm{d}x_i)$. If $\rho^{(1)}$ and $\rho^{(2)}$ exist,
we may consider the following second order summary statistic of
$\mathbf N$ (called pair correlation function)
\[
g(x_1,x_2)=\frac{\rho^{(2)}(x_1,x_2)}{\rho^{(1)}(x_1)\rho^{(1)}(x_2)}  \qquad \text{ for } \rho^{(1)}(x_1)>0, \rho^{(1)}(x_2)>0
\]
$g(x_1,x_2)=0$    when either $  \rho^{(1)}(x_1)=0$ or  $\rho^{(1)}(x_2)=0$.

Due to the interpretation of the joint intensities, if $g\leq 1$
$\lambda^{\otimes 2}$-a.e. then the points of $\mathbf N$ repel each
other (indeed the process is negative correlated and has an
anti-clumping behavior.)

$\mathbf N$ is said to be a determinantal process on $S$ with kernel
$K:S\times S\to\C$ and reference measure $\lambda$ if
\[
\rho^{(k)}(x_1,\ldots,x_k)=\mathrm{det}(K(x_i,x_j))_{1\leq i,j\leq
k},
\]
where $\mathrm{det}(K(x_i,x_j))_{1\leq i,j\leq k}$ is the
determinant of the $k\times k$-matrix with $ij$-entries
$K(x_i,x_j)$. From now on, we assume that $K$ is locally square
integrable on $S\times S$ with respect to $\lambda^{\otimes 2}$
and let
\[
\mathcal{K}f(x)=\int_{S}K(x,y)f(y)\,\lambda(\mathrm{d}y),\qquad
f\in L^2(S,\lambda).
\]
be the integral operator with kernel $K$ and reference measure
$\lambda$. Here $L^2(S,\lambda)$ is the space of functions
$f:S\to\mathbb C$ which are square integrable with respect to
$\lambda$. In the sequel, for a compact set $\Lambda'\subset S$, we
denote by $\mathcal{K}_{\Lambda'}$ the restriction of $\mathcal K$
to $\Lambda'$. If the operator $\mathcal{K}_{\Lambda'}$ is positive,
we denote by $\mathrm{Tr}(\mathcal{K}_{\Lambda'})$ the trace of
$\mathcal{K}_{\Lambda'}$.
To guarantee the existence and uniqueness (in law) of a
determinantal process with a given kernel $K$ and reference
measure $\lambda$ one assumes
\begin{itemize}
\item $\mathcal{K}$ is Hermitian, i.e.
$K(x_i,x_j)=\overline{K(x_j,x_i)}$, $\lambda^{\otimes 2}$-a.e.
\item  The spectrum of $\mathcal K$ is contained in $[0,1]$.
\item $\mathcal K$ is locally of trace class, i.e.
$\mathrm{Tr}(\mathcal{K}_{\Lambda'})<\infty$ for any compact
$\Lambda'\subset S$.
\end{itemize}
By the spectral theorem for compact and Hermitian operators, under
the above assumptions, for any fixed compact $\Lambda'\subset S$,
there exists an orthonormal basis $\{\varphi_{n,\Lambda'}\}_{n\geq
1}$ of $L^2(\Lambda',\lambda)$ of eigenfunctions of
$\mathcal{K}_{\Lambda'}$. We denote by $\{\kappa_n(\Lambda')\}_{n\geq
1}$ the corresponding eigenvalues, i.e.
$\mathcal{K}_{\Lambda'}\varphi_{n,\Lambda'}=\kappa_n(\Lambda')\varphi_{n,\Lambda'}$,
$n\geq 1$. Note that $\kappa_n(\Lambda')\in [0,1]$ for any $n\geq
1$, because the spectrum of $\mathcal K$ is contained in $[0,1]$.
Note also that the above conditions imply $K(x,x)\geq 0$,
$\lambda$-a.e..

We remark that for a determinantal process $\mathbf N$ on
$S$ with kernel $K$ and reference measure $\lambda$ we have
\begin{align}
g(x_1,x_2)&=\frac{K(x_1,x_1)K(x_2,x_2)-K(x_1,x_2)K(x_2,x_1)}{K(x_1,x_1)K(x_2,x_2)}\nonumber\\
&=1-\frac{K(x_1,x_2)K(x_2,x_1)}{K(x_1,x_1)K(x_2,x_2)}\nonumber\\
&=1-\frac{|K(x_1,x_2)|^2}{K(x_1,x_1)K(x_2,x_2)}\leq
1,\quad\text{$\lambda^{\otimes 2}$-a.e.}\label{eq:repel}
\end{align}
which shows the repulsiveness of determinantal processes. Here, in
\eqref{eq:repel} one uses first the Hermitianity of $\mathcal K$
and second that $K(x,x)\geq 0$ $\lambda$-a.e..

\begin{figure}[t!]
  \centering	
  \begin{subfigure}[t!]{0.33\textwidth}
    \centering
    \hspace*{-7mm}
   \includegraphics[height=40 mm]{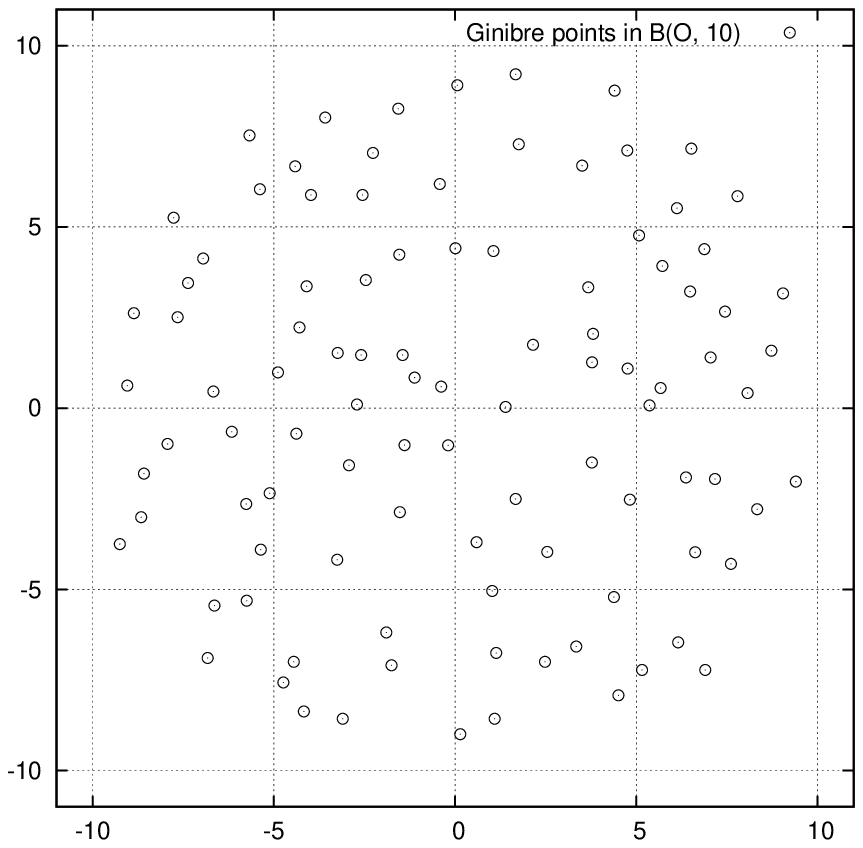}
    \caption{Ginibre}
    \label{fig:determintal}
  \end{subfigure}%
  \centering	
  \begin{subfigure}[t!]{0.33\textwidth}
    \centering
   \includegraphics[height=40 mm]{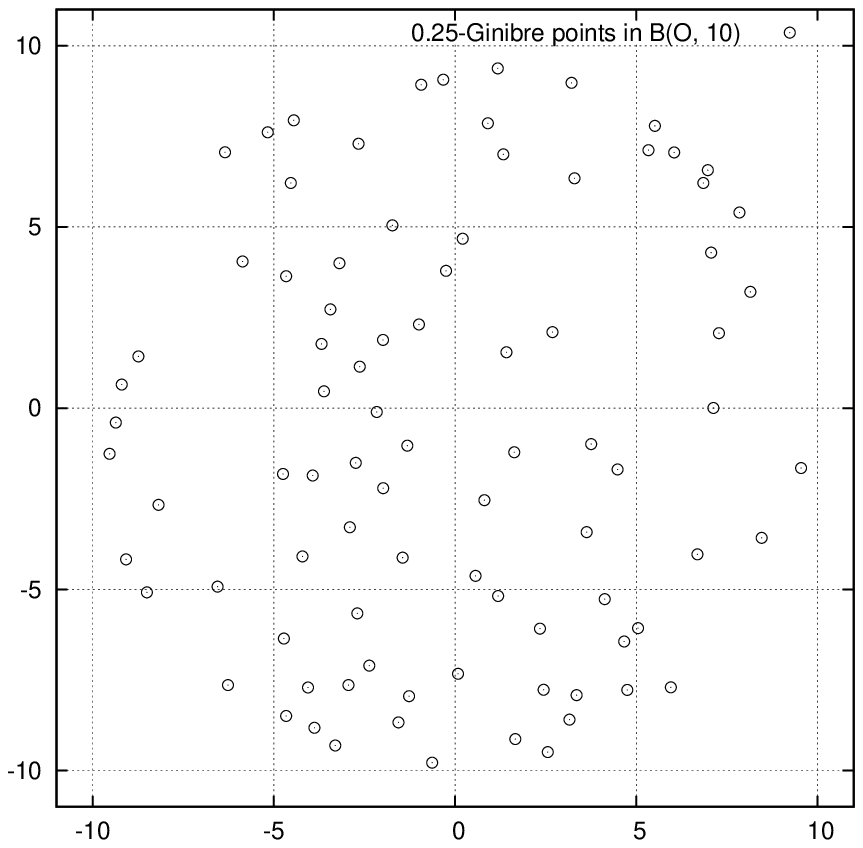}
    \caption{$\beta$-Ginibre}
    \label{fig:betadetermintal}
  \end{subfigure}%
  \begin{subfigure}[t!]{0.33\textwidth}
    \centering
    \hspace*{7mm}
    \includegraphics[height=40 mm]{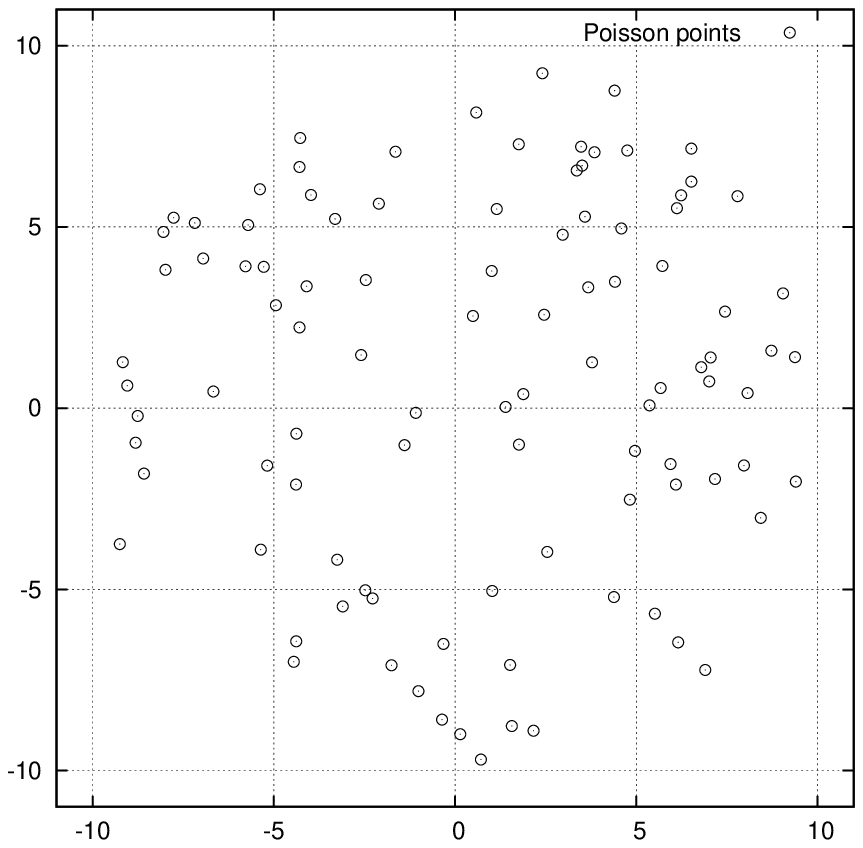}
    \caption{Poisson}
    \label{fig:poisson}
  \end{subfigure}
\caption{Realizations of the Ginibre process, the $\beta$-Ginibre process with $\beta=0.25$ and the homogeneous
Poisson process of intensity $1/\pi$ within the ball $b(O,10)$.}
  ~ 
\end{figure}

In this paper, we shall consider the Ginibre and more generally the $\beta$-Ginibre process.
The Ginibre process is a determinantal process on $S=\C$
with kernel $K$ and reference measure $\lambda$ defined
respectively by
\[
K(x,y)=\mathrm{e}^{x\overline{y}}\qquad\text{and}\qquad
\lambda(\mathrm{d}x)=\frac{1}{\pi}\mathrm{e}^{-|x|^2}\,\mathrm{d}x.
\]
Here $\mathrm{d}x$ denotes the Lebesgue measure on $\C$. The $\beta$-Ginibre process, $0<\beta\leq 1$,
is the point process obtained by retaining, independently and with probability $\beta$, each point of the Ginibre
process and then scaling by $\sqrt\beta$ the remaining points. Note that
the $1$-Ginibre process is the Ginibre process and that the $\beta$-Ginibre process converges weakly to
the homogeneous Poisson process of intensity $1/\pi$, as $\beta\to 0$ (this latter fact may be easily checked
proving that the Laplace functional of the $\beta$-Ginibre process converges to the Laplace functional
of the Poisson process of intensity $1/\pi$, as $\beta\to 0$; see e.g. Theorem 4 in \cite{decreusefond}.) In other
words the $\beta$-Ginibre processes, $0<\beta<1$, constitute an intermediate class between the homogeneous Poisson
process of intensity $1/\pi$ and the Ginibre process. We remark that the $\beta$-Ginibre processes, $0<\beta\leq 1$,
are still determinantal processes and
satisfy the usual conditions of existence and uniqueness (see e.g. \cite{goldman}.)
Figures~\ref{fig:determintal} and \ref{fig:betadetermintal} show, respectively, a realization of
the Ginibre processs and of the $\beta$-Ginibre process with $\beta=0.25$ within the ball $b(O,10)$.
For comparison, a realization of the homogeneous Poisson process of intensity $1/\pi$ within the ball $b(O,10)$
is reported in the Figure~\ref{fig:poisson}. Note that the points of the Ginibre process exhibit the highest degree of regularity,
while the points of the Poisson process exhibit the lowest degree of regularity.

\section{Statistical assumptions}\label{sec:ginibre}

Throughout this paper we assume that the signal power  is attenuated
according to the ideal Hertzian law, i.e.
\begin{equation*}
L(x)=\max\{R,|x|\}^{-\alpha},\quad\text{$R>0$,
$\alpha>2$.}
\end{equation*}
We recall that the simple point process $\bold N=\{Y_i\}_{i\geq 1}$ denotes
the locations of the nodes and $\{X_i\}_{i\geq 1}$ are the points of the reduced Palm version at the origin of $\bold N$,
i.e. $\bold N\setminus\{O\}\,|\, O\in\bold N$. In the following, any time we refer to a determinantal process
we identify the plane with $\mathbb{C}$.

\begin{Lemma}\label{le:idlaw}
Let $\{X_i\}_{i\geq 1}$ be a reduced Palm version at the origin
of a $\beta$-Ginibre process, $\{V_i\}_{i\geq 1}$ a Ginibre process
and $G$ a centered complex Gaussian random variable with $\EE[|G|^2]=1$.
The point process which is obtained by an independent thinning of
$\{\sqrt\beta V_i\}_{i\geq 1}$ with retention probability $\beta$ has the same law of the point
process which is obtained by adding to $\{X_i\}_{i\geq 1}$ the point $\sqrt{\beta} G$ with probability $\beta$.
\end{Lemma}
Given a measurable and bounded subset $\Lambda'$ of the plane, we
denote by $N(\Lambda')$ the number of points $\{X_i\}_{i\geq 1}$ in
$\Lambda'$.
\begin{Lemma}\label{le:ginibre}
\noindent$(i)$ Let $\{V_i\}_{i\geq 1}$ be a Ginibre process and $\{A_i\}_{i\geq 1}$ a sequence of independent
and identically distributed events, independent of $\{V_i\}_{i\geq 1}$. For any fixed $r\in (0,\infty)$ and $x_0\in\C$,
\begin{equation}\label{eq:krishnapurtinnato}
\PP\left(\sum_{i\geq 1}\ind_{b(x_0,r)}(V_i)\ind_{A_i}\geq
m\right)=\mathrm{e}^{-\frac{1}{2}m^2\log m(1+o(1))}, \quad\text{as $m\uparrow\infty$.}
\end{equation}
\noindent$(ii)$Let $\{X_i\}_{i\geq 1}$ be a reduced Palm version at the origin
of a $\beta$-Ginibre process. For any fixed $r\in (0,\infty)$ and $x_0\in\C$,
\[
\PP(N(b(x_0,r))\geq m)=\mathrm{e}^{-\frac{1}{2}m^2\log m(1+o(1))},\quad\text{as $m\uparrow\infty$.}
\]
\end{Lemma}
\begin{Lemma}\label{le:laplacemean}
Let $\{X_i\}_{i\geq 1}$ be a reduced Palm version at the origin
of a $\beta$-Ginibre process. For any compact $\Lambda'\subset\C$,
\begin{equation}\label{eq:trace}
\EE[N(\Lambda')]\leq\sum_{n\geq
1}\kappa_n(\Lambda'/\sqrt\beta)
<\infty
\end{equation}
and
\begin{equation}\label{eq:laplace}
\EE[\mathrm{e}^{\theta N(\Lambda')}]\leq\prod_{n\geq
1}(1+(\mathrm{e}^{\theta}-1)\kappa_n(\Lambda'/\sqrt\beta))<\infty,\qquad\theta\geq
0.
\end{equation}
Here
\[
\Lambda'/\beta=\{x\in\mathbb C:\,\,x=y/\sqrt\beta\,\,\text{for some}\,\,y\in\Lambda'\}
\]
and $\kappa_n(\Lambda'/\beta)$ are the eigenvalues of the integral operator, restricted to $\Lambda'/\beta$,
of the $1$-Ginibre process.
\end{Lemma}
Lemma \ref{le:idlaw} is a straightforward consequence of
Remark 24 in \cite{goldman} (see Theorem 1 in \cite{goldman} for the case $\beta=1$.)
The proofs of Lemmas \ref{le:ginibre} and \ref{le:laplacemean} are given in the Appendix.
Lemmas \ref{le:idlaw}, \ref{le:ginibre} and \ref{le:laplacemean} will come in handy in Sections~\ref{sec:ldptrunbsfad} and \ref{sec:ldptruncWeibfad}.

In Section \ref{sec:ldpexp}, we consider a general simple point process $\mathbf N$ on the plane
satisfying one of the following two light-tail conditions:
\begin{itemize}
\item when the fading is exponentially distributed (see Subsection \ref{par:exp}) we assume that
\begin{equation}\label{eq:superexpN}
\text{$\EE[\mathrm{e}^{\theta N(\Lambda)}]<\infty$ for any $\theta>0$;}
\end{equation}
\item when the fading is subexponential (see Subsection \ref{par:subexp}) we assume that
\begin{equation}\label{eq:lightN}
\text{$\exists$ $a>0$ such that
$\EE[\mathrm{e}^{\theta N(\Lambda)}]<\infty$ $\forall$ $\theta<a$.}
\end{equation}
\end{itemize}
Note that Conditions \eqref{eq:superexpN} and \eqref{eq:lightN}
are fairly general. The homogeneous Poisson process and the $\beta$-Ginibre process, $0<\beta\leq 1$,
represent just two particular point processes satisfying \eqref{eq:superexpN}, and therefore \eqref{eq:lightN}.
This is a simple consequence of the Slivnyak Theorem and Lemma
\ref{le:laplacemean}.

\section{Large deviations of the interference: bounded fading}\label{sec:ldptrunbsfad}

The standing assumptions of this section are: $\mathbf{N}$ is the
$\beta$-Ginibre process, $0<\beta\leq 1$; the fading random variables $Z_i$, $i\geq 1$, have bounded
support with supremum $B>0$.

\begin{Theorem}\label{thm:ldbd}
Under the foregoing assumptions, the family of random variables
$\{\varepsilon I_\Lambda\}_{\varepsilon>0}$ obeys an LDP on
$[0,\infty)$ with speed
$\frac{1}{\varepsilon^2}\log\frac{1}{\varepsilon}$ and good rate
function $I_1(x)=\frac{R^{2\alpha}x^2}{2 B^2}$.
\end{Theorem}

The proof of this theorem is based on the following lemmas whose
proofs are given below.
\begin{Lemma}\label{le:upperbdsupset}
Under the foregoing assumptions, for any $x\geq 0$,
\[
\limsup_{\varepsilon\to
0}\frac{\varepsilon^2}{\log(1/\varepsilon)}\log\PP(\varepsilon
I_\Lambda\geq x)\leq -I_1(x).
\]
\end{Lemma}
\begin{Lemma}\label{le:lowerbd}
Under the foregoing assumptions, for any $x\geq 0$,
\[
\liminf_{\varepsilon\to
0}\frac{\varepsilon^2}{\log(1/\varepsilon)}\log\PP(\varepsilon
I_\Lambda>x)\geq -I_1(x).
\]
\end{Lemma}

\noindent$\it{Proof\,\,of\,\,Theorem\,\,\ref{thm:ldbd}}$ The claim
follows by Proposition \ref{pro:crit} and Lemmas
\ref{le:upperbdsupset} and \ref{le:lowerbd}.
\\
\noindent$\square$
\\

\noindent$\it{Proof\,\,of\,\,Lemma\,\,\ref{le:upperbdsupset}}$ The
claim is clearly true if $x=0$. We prove the claim when $x>0$.
Since $\max\{R,|X_i-y|\}\geq R$ we have $L(X_i-y)\leq R^{-\alpha}$, $i\geq 1$, and so
\begin{equation}\label{eq:rev10ter}
\PP\left(\varepsilon I_\Lambda\geq x\right)\leq\PP\left(R^{-\alpha}\varepsilon\sum_{i\geq 1}Z_i\ind_{\Lambda}(X_i)\geq x\right),
\quad\text{$\varepsilon>0$}
\end{equation}
(it is worthwhile to remark that due to its generality
this bound will be used later on even to derive large deviation upper bounds
in the case of signals not necessarily bounded and nodes not necessarily distributed
as the reduced Palm version at the origin of a $\beta$-Ginibre process.)
Since $\Lambda$ is bounded and $y\in\Lambda^\circ$
there exists $\widetilde{R}>0$ so that $b(y,\widetilde{R})\supseteq\Lambda$. Combining this with \eqref{eq:rev10ter}
and the assumption on the support of the signals, for any $\varepsilon>0$, we have
\begin{align}
\PP(\varepsilon I_{\Lambda}\geq x)\leq \PP\left(\sum_{i\geq
1}\ind_{b(y,\widetilde{R})}(X_i)\geq \frac{R^\alpha
x}{B\varepsilon}\right)=\PP\left(N(b(y,\widetilde{R}))\geq\frac{R^\alpha
x}{B\varepsilon}\right).\label{eq:revrev3}
\end{align}
By this inequality and Lemma \ref{le:ginibre}$(ii)$ we then have
\begin{align}
\limsup_{\varepsilon\to
0}\frac{\varepsilon^2}{\log(1/\varepsilon)}\log\PP(\varepsilon
I_\Lambda\geq x)&\leq\limsup_{\varepsilon\to
0}\frac{\varepsilon^2}{\log(1/\varepsilon)}\log\PP\left(N(b(y,\widetilde{R}))\geq\frac{R^\alpha
x}{B\varepsilon}\right)\nonumber\\
&=-\frac{R^{2\alpha}x^2}{2 B^2},\nonumber
\end{align}
and the proof is completed (note that in the latter equality one makes use of the elementary
relation $\lim_{\varepsilon\to 0}\frac{\log(c/\varepsilon)}{\log(1/\varepsilon)}=1$, for any positive constant $c>0$.)
\\
\noindent$\square$
\\

\noindent$\it{Proof\,\,of\,\,Lemma\,\,\ref{le:lowerbd}}$ The idea is to produce a suitable lower bound
for the quantity $\PP(\varepsilon I_\Lambda> x)$ by a thinning argument. For this we shall
combine Lemma \ref{le:idlaw} and Lemma \ref{le:ginibre}$(i)$.
The claim of the lemma is clearly true if $x=0$ and so we consider $x>0$. Since
$y\in\Lambda^\circ$, there exists $r\in (0,R)$
such that $b(y,r)^\circ\subset\Lambda$. So,
for any $\varepsilon>0$, we have
\begin{align}
\hspace{-5 mm} \PP(\varepsilon I_\Lambda> x)&\geq\PP(\varepsilon I_{b(y,r)^\circ}>x)
=\PP\left(\sum_{i\geq 1}Z_i\ind_{b(y,r)^\circ}(X_i)>\frac{R^\alpha x}{\varepsilon}\right),\label{eq:rev8}
\end{align}
where the equality is a consequence of the fact that $r\in (0,R)$. Letting $\{U\}\cup\{U_i\}_{i\geq 1}$ denote a sequence of independent
random variables uniformly distributed on $[0,1]$ and
$Z$ denote a random variable distributed as $Z_1$, and assuming that the random variables
$\{U,Z\}\cup\{U_i\}_{i\geq 1}$ are independent of all the other
random quantities, we have
\begin{align}
&\PP\left(\sum_{i\geq 1}Z_i\ind_{b(y,r)^\circ}(X_i)>\frac{R^\alpha x}{\varepsilon}\right)\nonumber\\
&=\PP\left(\sum_{i\geq 1}Z_i\ind_{b(y,r)^\circ}(X_i)+Z\ind_{b(y,r)^\circ}(\sqrt\beta G)\ind\{U<\beta\}>\frac{R^\alpha x}{\varepsilon}+
Z\ind_{b(y,r)^\circ}(\sqrt\beta G)\ind\{U<\beta\}\right)\nonumber\\
&\geq\PP\left(\sum_{i\geq 1}Z_i\ind_{b(y,r)^\circ}(X_i)+Z\ind_{b(y,r)^\circ}(\sqrt\beta G)\ind\{U<\beta\}>\frac{R^\alpha x}{\varepsilon}+
B\right)\label{eq:revrev1}\\
&=\PP\left(\sum_{i\geq 1}Z_i\ind_{b(y,r)^\circ}(\sqrt\beta V_i)\ind\{U_i<\beta\}>\frac{R^\alpha x}{\varepsilon}+
B\right),\label{eq:rev9}
\end{align}
where \eqref{eq:revrev1} follows by the upper bound
\[
Z\ind_{b(y,r)^\circ}(\sqrt\beta G)\ind\{U<\beta\}\leq B
\]
and \eqref{eq:rev9} is consequence of Lemma \ref{le:idlaw}.
Since $Z_1$ has bounded support with supremum $B>0$, for arbitrarily small $\delta\in
(0,1)$ there exists $p_\delta>0$ such that
$\PP(Z_1>(1-\delta)B)=p_\delta$. Using the elementary relations
\[
1\geq\ind_{((1-\delta)B,\infty)}(Z_i),\quad
\ind_{b(y,r)^\circ}(\sqrt\beta V_i)=\ind_{b(y/\sqrt\beta,r/\sqrt\beta)^\circ}(V_i)
\]
we have
\begin{align}
&\PP\left(\sum_{i\geq 1}Z_i\ind_{b(y,r)^\circ}(\sqrt\beta V_i)\ind\{U_i<\beta\}>\frac{R^\alpha x}{\varepsilon}+B\right)\nonumber\\
&\geq\PP\left(\sum_{i\geq 1}Z_i\ind_{b(y/\sqrt\beta,r/\sqrt\beta)^\circ}(V_i)\ind\{U_i<\beta\}
\ind_{((1-\delta)B,\infty)}(Z_i)>\frac{R^\alpha x}{\varepsilon}+B\right).\label{eq:rev10}
\end{align}
Note also that
\begin{align}
&\PP\left(\sum_{i\geq 1}Z_i\ind_{b(y/\sqrt\beta,r/\sqrt\beta)^\circ}(V_i)\ind\{U_i<\beta\}
\ind_{((1-\delta)B,\infty)}(Z_i)>\frac{R^\alpha x}{\varepsilon}+B\right)\nonumber\\
&\geq\PP\left((1-\delta)B\sum_{i\geq 1}\ind_{b(y/\sqrt\beta,r/\sqrt\beta)^\circ}(V_i)\ind\{U_i<\beta\}
\ind_{((1-\delta)B,\infty)}(Z_i)>\frac{R^\alpha x}{\varepsilon}+B\right)\nonumber\\
&=\PP\left(\sum_{i\geq
1}\ind_{b(y/\sqrt\beta,r/\sqrt\beta)^\circ}(V_i)\ind\{U_i<\beta\}\ind_{((1-\delta)B,\infty)}(Z_i)>
\frac{R^\alpha x}{(1-\delta)B\varepsilon}+\frac{1}{1-\delta}\right)\nonumber\\
&\geq\PP\left(\sum_{i\geq
1}\ind_{b(y/\sqrt\beta,r/\sqrt\beta)^\circ}(V_i)\ind\{U_i<\beta\}\ind_{((1-\delta)B,\infty)}(Z_i)>
\left[\frac{R^\alpha x}{(1-\delta)B\varepsilon}+\frac{1}{1-\delta}\right]+1\right),\label{eq:dis1}
\end{align}
where the latter inequality follows by the definition of $[x]$ (i.e. the biggest integer
not exceeding $x$.) Collecting \eqref{eq:rev8}, \eqref{eq:rev9}, \eqref{eq:rev10} and \eqref{eq:dis1}
we deduce
\begin{align}
&\PP(\varepsilon I_\Lambda> x)\nonumber\\
&\geq\PP\left(\sum_{i\geq
1}\ind_{b(y/\sqrt\beta,r/\sqrt\beta)^\circ}(V_i)\ind\{U_i<\beta\}\ind_{((1-\delta)B,\infty)}(Z_i)>
\left[\frac{R^\alpha x}{(1-\delta)B\varepsilon}+\frac{1}{1-\delta}\right]+1\right).\label{eq:revrev2}
\end{align}
By this inequality and \eqref{eq:krishnapurtinnato}, we have
\begin{align}
\liminf_{\varepsilon\to
0}\frac{\varepsilon^2}{\log(1/\varepsilon)}\log\PP(\varepsilon
I_\Lambda>x)&\geq -\frac{1}{2}\lim_{\varepsilon\to
0}\frac{\varepsilon^2}{\log(1/\varepsilon)}\left[\frac{R^\alpha
x}{(1-\delta)B\varepsilon}\right]^2\log\left[\frac{R^\alpha
x}{(1-\delta)B\varepsilon}\right]
\nonumber\\
&=-\frac{1}{2}\frac{R^{2\alpha}x^2}{(1-\delta)^2 B^2}.\nonumber
\end{align}
The claim follows letting $\delta$ tend to zero.
\\
\noindent$\square$

We conclude this section stating the following immediate corollary
of Theorem \ref{thm:ldbd}.
\begin{Corollary}\label{cor:inttruncbd}
Under the assumptions of Theorem \ref{thm:ldbd},
\begin{equation}\label{resbounded}
\lim_{x\to\infty}\frac{\log\PP(I_\Lambda\geq x)}{x^2\log
x}=-\frac{1}{2}\frac{R^{2\alpha}}{B^2}.
\end{equation}
\end{Corollary}
The proof of Theorem \ref{thm:ldbd} suggests that large values of the interference are typically obtained as the sum
of the signals coming from a large number of interfering nodes. This interpretation follows by inequalities
\eqref{eq:revrev3} and \eqref{eq:revrev2}.

Now we can compare  \eqref{resbounded} against  its analogue for Poisson networks derived
in  \cite{ganesh} and here repeated (see also Proposition 5.1 in \cite{leonardi}):
\begin{Proposition}
If $\mathbf N$ is a homogeneous Poisson process and the
fading random variables have bounded
support with supremum $B>0$,
\[
\lim_{x\to\infty}\frac{\log\PP(I_\Lambda\geq x)}{x\log
x}=-\frac{R^{\alpha}}{B}.
\]
\end{Proposition}
We conclude that: $i)$ on the log-scale, the asymptotic behavior of the tail of the interference
is insensitive to the choice of the particular $\beta$-Ginibre network model
(it does not depend on $0<\beta\leq 1$), as a consequence of the fact that
the tail of the number of points falling in a ball has the same asymptotic behavior for any value of
$\beta\in (0,1]$ (see Lemma \ref{le:ginibre});
$ii)$ the tail of the interference in the $\beta$-Ginibre network model is significantly lighter than the tail of the interference
in the Poisson network model. This is a direct consequence of the repulsiveness of the $\beta$-Ginibre process, $0<\beta\leq 1$.
Since the $\beta$-Ginibre process converges weakly to the homogeneous Poisson process with intensity $1/\pi$, as $\beta \to 0$,
the tail of the interference exhibits a discontinuous behavior with respect to the convergence in law.

>From an application point of view, our results lead to the following conclusion:
when transmissions are marginally affected by fading such as in outdoor scenarios with (almost) line of sight transmissions,
the impact of the node placement can be significant. Network planners should place network nodes as regularly as possible,
avoiding concentration of nodes in small areas.

\section{Large deviations of the interference: Weibull superexponential fading}\label{sec:ldptruncWeibfad}

The standing assumptions of this section are: $\mathbf N$ is the
$\beta$-Ginibre process, $0<\beta\leq 1$; the fading random variables $Z_i$, $i\geq 1$, are Weibull
superexponential in the sense that $-\log\PP(Z_1>z)\sim
cz^\gamma$, for some constants $c>0$ and $\gamma>1$.

Hereafter, for a constant $\mu\in\R$ and $x>0$ we use the standard notation $\log^{\mu}x=(\log x)^\mu$.

\begin{Theorem}\label{thm:ldW}
Under the foregoing assumptions, the family of random variables
$\{\varepsilon I_\Lambda\}_{\varepsilon>0}$ obeys an LDP on
$[0,\infty)$ with speed
$\frac{1}{\varepsilon^{2\gamma/(\gamma+1)}}\log^{(\gamma-1)/(\gamma+1)}\left(\frac{1}{\varepsilon}\right)$
and good rate function
$$I_2(x)=\frac{1}{2}R^{2\alpha\gamma/(\gamma+1)}\left(\frac{\gamma}{\gamma-1}\right)^{(\gamma-1)/(\gamma+1)}(c(\gamma+1))^{2/(\gamma+1)}x^{2\gamma/(\gamma+1)}.$$
\end{Theorem}

The proof of this theorem is based on the following lemmas whose
proofs are given below.
\begin{Lemma}\label{le:upperWsupset}
Under the foregoing assumptions, for any $x\geq 0$,
\begin{equation*}
\limsup_{\varepsilon\to
0}\frac{\varepsilon^{2\gamma/(\gamma+1)}}{\log^{(\gamma-1)/(\gamma+1)}(1/\varepsilon)}
\log\PP(\varepsilon I_\Lambda\geq x)\leq -I_2(x).
\end{equation*}
\end{Lemma}
\begin{Lemma}\label{le:lowerW}
Under the foregoing assumptions, for any $x\geq 0$,
\begin{equation*}
\liminf_{\varepsilon\to
0}\frac{\varepsilon^{2\gamma/(\gamma+1)}}{\log^{(\gamma-1)/(\gamma+1)}(1/\varepsilon)}
\log\PP(\varepsilon I_\Lambda>x)\geq -I_2(x).
\end{equation*}
\end{Lemma}

\noindent$\it{Proof\,\,of\,\,Theorem\,\,\ref{thm:ldW}}$ The claim
follows by Proposition \ref{pro:crit} and Lemmas
\ref{le:upperWsupset} and \ref{le:lowerW}.
\\
\noindent$\square$
\\

\noindent$\it{Proof\,\,of\,\,Lemma\,\,\ref{le:upperWsupset}}$ The
claim is clearly true if $x=0$. We prove the claim when $x>0$ in four steps.
In the first step we provide a general upper bound for $\PP(\varepsilon I_\Lambda\geq x)$, $\varepsilon>0$,
by applying the Chernoff bound (it is worthwhile to remark that due to its generality
the bound obtained in this step will be used later on even to derive large deviation upper bounds in the case of exponential
signals and nodes not necessarily distributed as the reduced Palm version at the origin
of a $\beta$-Ginibre process.) In the second step, using the determinantal structure of the Ginibre process and the bound derived in Step 1, we give
a further upper bound for $\PP(\varepsilon I_\Lambda\geq x)$. In the third step we show how the conclusion can be derived by the bound
proved in Step 2. This is done up to a technical point which is addressed in the subsequent Step 4.
\\
\noindent${\it Step\,\,1:\,\,An\,\,upper\,\,bound\,\,for\,\,\PP(\varepsilon I_\Lambda\geq x).}$
Let $\Lambda'$ be a bounded set of the complex plane such that $\Lambda'\supseteq\Lambda$ and let
$\theta>0$ be an arbitrary positive constant. By the Chernoff bound and the independence, we deduce
\begin{align}
\PP\left(\varepsilon R^{-\alpha}\sum_{i=1}^{N(\Lambda')}Z_i\geq
x\right)&\leq\exp\left(-\theta
x+\log\EE\left[\mathrm{e}^{\theta\varepsilon
R^{-\alpha}\sum_{i=1}^{N(\Lambda')}Z_i}\right]\right)\nonumber\\
&=\exp\left(-\theta
x+\log\EE\left[\EE\left[\mathrm{e}^{\theta\varepsilon
R^{-\alpha}Z_1}\right]^{N(\Lambda')}\right]\right).\label{eqchernoff}
\end{align}
Combining \eqref{eq:rev10ter} and \eqref{eqchernoff}, we deduce
\begin{equation}\label{eq:rev10bis}
\PP\left(\varepsilon I_\Lambda\geq x\right)\leq\exp\left(-\theta
x+\log\EE\left[\EE\left[\mathrm{e}^{\theta\varepsilon
R^{-\alpha}Z_1}\right]^{N(\Lambda')}\right]\right)
\end{equation}
(note that by the assumption on the distribution of $Z_1$ one has
$\EE\left[\mathrm{e}^{\delta Z_1}\right]<\infty$ for any
$\delta>0$ and so the bound is finite.)
\\
\noindent${\it Step\,\,2:\,\,A\,\,further\,\,upper\,\,bound\,\,for\,\,\PP(\varepsilon I_\Lambda\geq x).}$
Let $\widetilde{R}>0$ be such that $b(O,\widetilde{R})\supseteq\Lambda$ and set $R'=\widetilde{R}/\sqrt\beta$.
Using \eqref{eq:laplace} we deduce
\begin{align}
\log\EE\left[\EE\left[\mathrm{e}^{\theta\varepsilon
R^{-\alpha}Z_1}\right]^{N(b(O,\widetilde{R}))}\right]
&\leq\log\prod_{n\geq
1}\left(1+\left(\EE\left[\mathrm{e}^{\theta\varepsilon
R^{-\alpha}Z_1}\right]-1\right)\kappa_n(b(O,R'))\right)\nonumber\\
&=\sum_{n\geq 1}\log\left(1+(\EE[\mathrm{e}^{\theta\varepsilon
R^{-\alpha} Z_1}]-1)\kappa_n(b(O,R'))\right).\label{eq:serieslog}
\end{align}
Combining \eqref{eq:rev10bis} with $\Lambda'=b(O,\widetilde{R})$ and \eqref{eq:serieslog}, for any $\varepsilon,x>0$,
we have
\begin{align}
\PP\left(\varepsilon I_\Lambda\geq x\right)&\leq
\exp\left(-\theta
x+\log\EE\left[\EE\left[\mathrm{e}^{\theta\varepsilon
R^{-\alpha}Z_1}\right]^{N(b(O,\widetilde{R}))}\right]\right)\nonumber\\
&\leq\exp\left(-\theta
x+
\sum_{n\geq 1}\log\left(1+(\EE[\mathrm{e}^{\theta\varepsilon
R^{-\alpha} Z_1}]-1)\kappa_n(b(O,R'))\right)
\right).\label{eq:rev11}
\end{align}
\noindent${\it Step\,\,3:\,\,Conclusion\,\,of\,\,the\,\,proof.}$
By \eqref{eq:rev11}, for any
$0<\varepsilon<\min\{1,x\}$, we have
\begin{align}
&\frac{\varepsilon^{2\gamma/(\gamma+1)}}{\log^{(\gamma-1)/(\gamma+1)}(1/\varepsilon)}\log\PP\left(\varepsilon I_\Lambda\geq
x\right)\nonumber\\
&\leq-\frac{\varepsilon^{2\gamma/(\gamma+1)}\theta
x}{\log^{(\gamma-1)/(\gamma+1)}(1/\varepsilon)}
+\frac{\varepsilon^{2\gamma/(\gamma+1)}}{\log^{(\gamma-1)/(\gamma+1)}(1/\varepsilon)}
\sum_{n\geq 1}\log\left(1+(\EE[\mathrm{e}^{\theta\varepsilon
R^{-\alpha} Z_1}]-1)\kappa_n(b(O,R'))\right).\label{eq:IIpezzo}
\end{align}
>From now on we take
\begin{equation*}
\theta=\frac{R^\alpha\tilde{\gamma}}{\varepsilon}\left(\frac{x}{\varepsilon}\log\frac{x}{\varepsilon}\right)^{(\gamma-1)/(\gamma+1)},
\end{equation*}
where
\begin{equation*}
\tilde\gamma=\frac{1}{2}\left(\frac{R^\alpha\gamma}{\gamma-1}\right)^{(\gamma-1)/(\gamma+1)}(c(\gamma+1))^{2/(\gamma+1)}.
\end{equation*}
Note that
\begin{equation}\label{eq:Ipezzo}
\lim_{\varepsilon\to
0}\frac{\varepsilon^{2\gamma/(\gamma+1)}\theta
x}{\log^{(\gamma-1)/(\gamma+1)}(1/\varepsilon)}=R^\alpha\tilde{\gamma}x^{2\gamma/(\gamma+1)}.
\end{equation}
We shall show in the next step that
\begin{equation}\label{eq:IIIpezzo}
\lim_{\varepsilon\to
0}\frac{\varepsilon^{2\gamma/(\gamma+1)}}{\log^{(\gamma-1)/(\gamma+1)}(1/\varepsilon)}
\sum_{n\geq 1}\log\left(1+(\EE[\mathrm{e}^{\theta\varepsilon
R^{-\alpha} Z_1}]-1)\kappa_n(b(O,R'))\right) =0.
\end{equation}
The claim follows taking the $\limsup$ as $\varepsilon\to 0$ in the inequality \eqref{eq:IIpezzo} and using \eqref{eq:Ipezzo}
and \eqref{eq:IIIpezzo}.
\\
\noindent${\it Step\,\,4:\,\,Proof\,\,of\,\,\eqref{eq:IIIpezzo}.}$
We start recalling that by Lemma 8 in \cite{ganesh} we have
\begin{equation}\label{eq:loglaplimW}
\lim_{\theta\to\infty}\frac{\log\EE[\mathrm{e}^{\theta
Z_1}]}{\gamma'\theta^{\gamma/(\gamma-1)}}=1,
\end{equation}
where
$\gamma'=(\gamma-1)\gamma^{-\gamma/(\gamma-1)}c^{-1/(\gamma-1)}$.
Since the eigenvalues $\kappa_n(b(O,R'))$ belong to $[0,1]$, by
\eqref{eq:loglaplimW} we deduce
\begin{align}
0&\leq\limsup_{\varepsilon\to
0}\frac{\varepsilon^{2\gamma/(\gamma+1)}}{\log^{(\gamma-1)/(\gamma+1)}(1/\varepsilon)}
\log\left(1+(\EE[\mathrm{e}^{\theta\varepsilon R^{-\alpha}
Z_1}]-1)\kappa_n(b(O,R'))\right)\nonumber\\
&\leq\limsup_{\varepsilon\to
0}\frac{\varepsilon^{2\gamma/(\gamma+1)}}{\log^{(\gamma-1)/(\gamma+1)}(1/\varepsilon)}
\log\EE[\mathrm{e}^{\theta\varepsilon R^{-\alpha}
Z_1}]\nonumber\\
&=\gamma'\tilde{\gamma}^{\gamma/(\gamma-1)}\lim_{\varepsilon\to
0}\frac{\varepsilon^{2\gamma/(\gamma+1)}}{\log^{(\gamma-1)/(\gamma+1)}(1/\varepsilon)}
\left(\frac{x}{\varepsilon}\log\frac{x}{\varepsilon}\right)^{\gamma/(\gamma+1)}=0.\nonumber
\end{align}
So, for \eqref{eq:IIIpezzo} we only need to check that we can
interchange the limit with the infinite sum. To this aim, we shall
prove that there exists a right neighborhood of zero, say
$\mathcal{N}_0$, such that
\[
\sum_{n\geq
1}\sup_{\varepsilon\in\mathcal{N}_0}\frac{\varepsilon^{2\gamma/(\gamma+1)}}{\log^{(\gamma-1)/(\gamma+1)}(1/\varepsilon)}\log\left(1+(\EE[\mathrm{e}^{\theta\varepsilon
R^{-\alpha} Z_1}]-1)\kappa_n(b(O,R'))\right)<\infty.
\]
By \eqref{eq:loglaplimW}, for any $\delta>0$ there
exists $\varepsilon_\delta\in (0,\min\{1,x\})$ such that for any
$\varepsilon\in (0,\varepsilon_\delta)$
\[
\EE[\mathrm{e}^{\theta\varepsilon
R^{-\alpha}Z_1}]\leq\exp\left(C_\delta\left(\frac{x}{\varepsilon}\log\frac{x}{\varepsilon}\right)^{\gamma/(\gamma+1)}\right)
\]
where
$C_\delta=(1+\delta)\gamma'\tilde{\gamma}^{\gamma/(\gamma-1)}$.
Therefore, for all $\varepsilon\in (0,\varepsilon_\delta)$, we
have
\begin{align}
&\frac{\varepsilon^{2\gamma/(\gamma+1)}}{\log^{(\gamma-1)/(\gamma+1)}(1/\varepsilon)}\log\left(1+(\EE[\mathrm{e}^{\theta\varepsilon
R^{-\alpha}
Z_1}]-1)\kappa_n(b(O,R'))\right)\nonumber\\
&\leq\frac{\varepsilon^{2\gamma/(\gamma+1)}}{\log^{(\gamma-1)/(\gamma+1)}(1/\varepsilon)}\log\left(1+\left(
\exp\left(C_\delta\left(\frac{x}{\varepsilon}\log\frac{x}{\varepsilon}\right)^{\gamma/(\gamma+1)}\right)
-1\right)\kappa_n(b(O,R'))\right).\label{eq:genter}
\end{align}
Consequently, it suffices to prove that there exists a right
neighborhood of zero contained in $(0,\varepsilon_\delta)$, say
$\mathcal{N}_0'$, such that
\begin{align}
\sum_{n\geq
1}\sup_{\varepsilon\in\mathcal{N}_0'}\frac{\varepsilon^{2\gamma/(\gamma+1)}}{\log^{(\gamma-1)/(\gamma+1)}(1/\varepsilon)}\log\left(1+\left(
\exp\left(C_\delta\left(\frac{x}{\varepsilon}\log\frac{x}{\varepsilon}\right)^{\gamma/(\gamma+1)}\right)
-1\right)\kappa_n(b(O,R'))\right)<\infty.\label{eq:series}
\end{align}
The first derivative (with respect to $\varepsilon$) of the term
in the right-hand side of \eqref{eq:genter} is equal to
\begin{align}
&\frac{2\gamma}{\gamma+1}\varepsilon^{(\gamma-1)/(\gamma+1)}\frac{\log\left(1+\left(\exp\left(C_\delta\left(\frac{x}{\varepsilon}\log\frac{x}{\varepsilon}\right)^{\gamma/(\gamma+1)}\right)
-1\right)\kappa_n(b(O,R'))\right)}{\log^{(\gamma-1)/(\gamma+1)}(1/\varepsilon)}\nonumber\\
&\,\,\,\,\,\,\,\,\,\,\,\,+\frac{\gamma-1}{\gamma+1}
\frac{\varepsilon^{(\gamma-1)/(\gamma+1)}\log\left(1+\left(\exp\left(C_\delta\left(\frac{x}{\varepsilon}\log\frac{x}{\varepsilon}\right)^{\gamma/(\gamma+1)}\right)-1\right)\kappa_n(b(O,R'))\right)}
{\log^{2\gamma/(\gamma+1)}(1/\varepsilon)}\nonumber\\
&\,\,\,\,\,\,\,\,\,\,\,\,-\frac{\frac{\gamma}{\gamma+1}x^{\gamma/(\gamma+1)}
C_\delta\exp\left(C_\delta\left(\frac{x}{\varepsilon}\log\frac{x}{\varepsilon}\right)^{\gamma/(\gamma+1)}\right)
\kappa_n(b(O,R'))}
{1+\left(\exp\left(C_\delta\left(\frac{x}{\varepsilon}\log\frac{x}{\varepsilon}\right)^{\gamma/(\gamma+1)}\right)-1\right)
\kappa_n(b(O,R'))}\nonumber\\
&\,\,\,\,\,\,\,\,\,\,\,\,\,\,\,\,\,\,\,\,\,\,\,\,
\,\,\,\,\,\,\,\,\,\,\,\,\,\,\,\,\,\,\,\,\,\,\,\,
\times\frac{1}{\varepsilon^{1/(\gamma+1)}}
\frac{\log^{\gamma/(\gamma+1)}(x/\varepsilon)}{\log^{(\gamma-1)/(\gamma+1)}(1/\varepsilon)}
\left(1+\frac{1}{\log^{\gamma/(\gamma+1)}(x/\varepsilon)\log^{1/(\gamma+1)}(1/\varepsilon)}\right)
.\nonumber
\end{align}
This quantity is bigger than or equal to zero if and only if
\begin{align}
&2\gamma\,\varepsilon^{\gamma/(\gamma+1)}\frac{\log\left(1+\left(\exp\left(C_\delta\left(\frac{x}{\varepsilon}\log\frac{x}{\varepsilon}\right)^{\gamma/(\gamma+1)}\right)
-1\right)\kappa_n(b(O,R'))\right)}{\log^{\gamma/(\gamma+1)}(1/\varepsilon)}\label{eq:H1}\\
&\,\,\,\,\,\,\,\,\,\,\,\,+(\gamma-1)\,
\frac{\varepsilon^{\gamma/(\gamma+1)}\log\left(1+\left(\exp\left(C_\delta\left(\frac{x}{\varepsilon}\log\frac{x}{\varepsilon}\right)^{\gamma/(\gamma+1)}\right)-1\right)\kappa_n(b(O,R'))\right)}
{\log^{(2\gamma+1)/(\gamma+1)}(1/\varepsilon)}\label{eq:H2}\\
&\,\,\,\,\,\,\,\,\,\,\,\,\,\,\,\,\,\,\,\,\,\,\,\,
\,\,\,\,\,\,\,\,\,\,\,\,\,\,\,\,\,\,\,\,\,\,\,\,
\geq\gamma\,\frac{x^{\gamma/(\gamma+1)}
C_\delta\exp\left(C_\delta\left(\frac{x}{\varepsilon}\log\frac{x}{\varepsilon}\right)^{\gamma/(\gamma+1)}\right)
\kappa_n(b(O,R'))}
{1+\left(\exp\left(C_\delta\left(\frac{x}{\varepsilon}\log\frac{x}{\varepsilon}\right)^{\gamma/(\gamma+1)}\right)-1\right)
\kappa_n(b(O,R'))}\nonumber\\
&\,\,\,\,\,\,\,\,\,\,\,\,\,\,\,\,\,\,\,\,\,\,\,\,
\,\,\,\,\,\,\,\,\,\,\,\,\,\,\,\,\,\,\,\,\,\,\,\,
\,\,\,\,\,\,\,\,\,\,\,\,\,\,\,\,\,\,\,\,\,\,\,\, \times
\frac{\log^{\gamma/(\gamma+1)}(x/\varepsilon)}{\log^{\gamma/(\gamma+1)}(1/\varepsilon)}
\left(1+\frac{1}{\log^{\gamma/(\gamma+1)}(x/\varepsilon)\log^{1/(\gamma+1)}(1/\varepsilon)}\right)\nonumber
\end{align}
Since $\kappa_n(b(O,R'))\in [0,1]$, we have
\begin{align}
&\gamma\,\frac{x^{\gamma/(\gamma+1)}C_\delta\exp\left(C_\delta
\left(\frac{x}{\varepsilon}\log\frac{x}{\varepsilon}\right)^{\gamma/(\gamma+1)}\right)\kappa_n(b(O,R'))}
{1+\left(\exp\left(C_\delta\left(\frac{x}{\varepsilon}\log\frac{x}{\varepsilon}\right)^{\gamma/(\gamma+1)}\right)-1\right)
\kappa_n(b(O,R'))}\nonumber\\
&\,\,\,\,\,\,\,\,\,\,\,\,\,\,\,\,\,\,\,\,\,\,\,\,
\,\,\,\,\,\,\,\,\,\,\,\,\,\,\,\,\,\,\,\,\,\,\,\,\times
\frac{\log^{\gamma/(\gamma+1)}(x/\varepsilon)}{\log^{\gamma/(\gamma+1)}(1/\varepsilon)}
\left(1+\frac{1}{\log^{\gamma/(\gamma+1)}(x/\varepsilon)\log^{1/(\gamma+1)}(1/\varepsilon)}\right)\nonumber
\\
&\leq J(\varepsilon):=\gamma\,x^{\gamma/(\gamma+1)}C_\delta
\frac{\log^{\gamma/(\gamma+1)}(x/\varepsilon)}{\log^{\gamma/(\gamma+1)}(1/\varepsilon)}
\left(1+\frac{1}{\log^{\gamma/(\gamma+1)}(x/\varepsilon)\log^{1/(\gamma+1)}(1/\varepsilon)}\right)
.\nonumber
\end{align}
Therefore, the first derivative of the term in the right-hand side
of \eqref{eq:genter} is bigger than or equal to zero if
\[
H^{(1)}(\varepsilon,\kappa_n(b(O,R')))+H^{(2)}(\varepsilon,\kappa_n(b(O,R')))\geq
J(\varepsilon),
\]
where, for ease of notation, we denoted by $H^{(1)}(\varepsilon,\kappa_n(b(O,R')))$ the term in
\eqref{eq:H1} and by $H^{(2)}(\varepsilon,\kappa_n(b(O,R')))$ the term in
\eqref{eq:H2}. By Remark 3.3 in \cite{shirai} we have
$\kappa_n(b(O,R'))=\PP(\mathrm{Po}(R'^2)\geq n+1)$, where
$\mathrm{Po}(R'^2)$ is a Poisson random variable with mean $R'^2$.
So the sequence $\{\kappa_n(b(O,R'))\}_{n\geq 1}$ is decreasing
(and decreases to zero.) Hence
\begin{align}
&\lim_{\varepsilon\to 0}\sup_{n\geq
1}(H^{(1)}(\varepsilon,\kappa_n(b(O,R')))+H^{(2)}(\varepsilon,\kappa_n(b(O,R'))))\nonumber\\
&\,\,\,\,\,\,=\lim_{\varepsilon\to
0}(H^{(1)}(\varepsilon,\kappa_1(b(O,R')))+H^{(2)}(\varepsilon,\kappa_1(b(O,R'))))\nonumber\\
&\,\,\,\,\,\,=2\gamma\,x^{\gamma/(\gamma+1)}C_\delta.\label{eq:lim1}
\end{align}
Furthermore,
\begin{equation}\label{eq:lim2}
\lim_{\varepsilon\to
0}J(\varepsilon)=\gamma\,x^{\gamma/(\gamma+1)}C_\delta.
\end{equation}
Let $\eta>0$ be such that
$\gamma\,x^{\gamma/(\gamma+1)}C_\delta>2\eta$. By \eqref{eq:lim1}
and \eqref{eq:lim2}, there exists $\varepsilon_\eta>0$ such that
for all
$0<\varepsilon<\min\{\varepsilon_\delta,\varepsilon_\eta\}$
\begin{align}
\sup_{n\geq
1}(H^{(1)}(\varepsilon,\kappa_n(b(O,R')))+H^{(2)}(\varepsilon,\kappa_n(b(O,R'))))&>
2\gamma\,x^{\gamma/(\gamma+1)}C_\delta-\eta\nonumber\\
&>\gamma\,x^{\gamma/(\gamma+1)}C_\delta+\eta>J(\varepsilon).\nonumber
\end{align}
This guarantees that the function of $\varepsilon$ in the
right-hand side of \eqref{eq:genter} is non-decreasing on
$(0,\min\{\varepsilon_\delta,\varepsilon_\eta\})$. Consequently,
setting
$\bar{\varepsilon}:=\min\{\varepsilon_\delta,\varepsilon_\eta\}$
and $\mathcal{N}_0'=(0,\overline{\varepsilon})$ we have
\begin{align}
&\sum_{n\geq
1}\sup_{\varepsilon\in\mathcal{N}_0'}\frac{\varepsilon^{2\gamma/(\gamma+1)}}{\log^{(\gamma-1)/(\gamma+1)}(1/\varepsilon)}
\log\left(1+(\EE[\mathrm{e}^{\theta\varepsilon R^{-\alpha}
Z_1}]-1)\kappa_n(b(O,R'))\right)\nonumber\\
&\leq\sum_{n\geq
1}\frac{\bar{\varepsilon}^{2\gamma/(\gamma+1)}}{\log^{(\gamma-1)/(\gamma+1)}(1/\bar{\varepsilon})}
\log\left(1+\left(
\exp\left(C_\delta\left(\frac{x}{\bar{\varepsilon}}\log\frac{x}{\bar{\varepsilon}}\right)^{\gamma/(\gamma+1)}\right)
-1\right)\kappa_n(b(O,R'))\right)\nonumber\\
&\leq\frac{\bar{\varepsilon}^{2\gamma/(\gamma+1)}}{\log^{(\gamma-1)/(\gamma+1)}(1/\bar{\varepsilon})}
\exp\left(C_\delta\left(\frac{x}{\bar{\varepsilon}}\log\frac{x}{\bar{\varepsilon}}\right)^{\gamma/(\gamma+1)}\right)
\sum_{n\geq 1}\kappa_n(b(O,R'))<\infty,\nonumber
\end{align}
where the latter inequality follows by
$\log(1+x)\leq x$, $x>-1$, and Lemma \ref{le:laplacemean}. The
proof is completed.
\\
\noindent$\square$
\\

\noindent$\it{Proof\,\,of\,\,Lemma\,\,\ref{le:lowerW}}$ Since the
claim is true if $x=0$, we take $x>0$. Since $y\in\Lambda^\circ$,
there exists $r\in (0,R)$ such that $b(y,r)^\circ\subset\Lambda$. For all
$\varepsilon>0$ and $n\geq 1$, we have
\begin{align}
\PP(\varepsilon I_\Lambda> x)\geq \PP(\varepsilon
I_{b(y,r)^\circ}>x)&
=\PP\left(\sum_{i\geq 1}Z_i\ind_{b(y,r)^\circ}(X_i)>\frac{R^\alpha x}{\varepsilon}\right)\nonumber\\
&\geq\PP\left(\sum_{i\geq 1}Z_i\ind_{b(y,r)^\circ}(X_i)>\frac{R^\alpha x}{\varepsilon}, N(b(y,r)^\circ)\geq n\right).\label{eq:rev12}
\end{align}
Define the event
\begin{equation}\label{eq:A}
A_\varepsilon^{(n)}:=\left\{\min\{Z_1,\ldots,Z_n\}>\frac{R^\alpha x}{n\varepsilon},N(b(y,r)^\circ)\geq n\right\}.
\end{equation}
Since
\begin{align}
&\PP\left(\sum_{i\geq 1}Z_i\ind_{b(y,r)^\circ}(X_i)>\frac{R^\alpha x}{\varepsilon}, N(b(y,r)^\circ)\geq n\right)\nonumber\\
&\geq\PP\left(\sum_{i=1}^{n}Z_i>\frac{R^\alpha x}{\varepsilon}, N(b(y,r)^\circ)\geq n\right)\nonumber\\
&\geq\PP\left(A_\varepsilon^{(n)}\right),\label{eq:rev13}
\end{align}
combining \eqref{eq:rev12} and \eqref{eq:rev13} and using the independence and that the signals are identically distributed, we have
\begin{align}
\PP(\varepsilon I_\Lambda> x)\geq\PP(A_\varepsilon^{(n)})=\PP(N(b(y,r)^\circ)\geq n)\PP\left(Z_1>\frac{R^\alpha
x}{n\varepsilon}\right)^n.\label{eq:weib0}
\end{align}
For $0<\varepsilon<1$, define the integer
\begin{equation}\label{eq:n}
n=\left[\frac{\kappa}{\varepsilon^{\gamma/(\gamma+1)}\log^{1/(\gamma+1)}(1/\varepsilon)}\right],
\end{equation}
where $\kappa>0$ is a constant which will be specified later. By
Lemma \ref{le:ginibre}$(ii)$, as $\varepsilon\to 0$, we deduce
\begin{align}
&-\log\PP\left(N(b(y,r)^\circ)\geq n\right)\nonumber\\
&\sim\frac{1}{2}\frac{\kappa^2}{\varepsilon^{2\gamma/(\gamma+1)}\log^{2/(\gamma+1)}(1/\varepsilon)}
\log\left(\frac{1}{\varepsilon^{\gamma/(\gamma+1)}\log^{1/(\gamma+1)}(1/\varepsilon)}\right)\nonumber\\
&\sim\frac{\gamma}{2(\gamma+1)}\frac{\kappa^2}{\varepsilon^{2\gamma/(\gamma+1)}}\log^{(\gamma-1)/(\gamma+1)}(1/\varepsilon).\label{eq:weib1}
\end{align}
Here, for the latter relation we used the following elementary computation
\begin{align}
&\frac{1}{\log^{2/(\gamma+1)}(1/\varepsilon)}
\log\left(\frac{1}{\varepsilon^{\gamma/(\gamma+1)}\log^{1/(\gamma+1)}(1/\varepsilon)}\right)\nonumber\\
&=\frac{\log(1/\varepsilon^{\gamma/(\gamma+1)})}{\log^{2/(\gamma+1)}(1/\varepsilon)}+
\frac{\log\left(1/\log^{1/(\gamma+1)}(1/\varepsilon)\right)}{\log^{2/(\gamma+1)}(1/\varepsilon)}\nonumber\\
&=\frac{\gamma}{\gamma+1}\frac{\log(1/\varepsilon)}{\log^{2/(\gamma+1)}(1/\varepsilon)}+
\frac{\log\left(1/\log^{1/(\gamma+1)}(1/\varepsilon)\right)}{\log^{2/(\gamma+1)}(1/\varepsilon)}\nonumber\\
&=\frac{\gamma}{\gamma+1}\log^{(\gamma-1)/(\gamma+1)}(1/\varepsilon)-\frac{1}{\gamma+1}
\frac{\log\log(1/\varepsilon)}{\log^{2/(\gamma+1)}(1/\varepsilon)}\nonumber\\
&\sim\frac{\gamma}{\gamma+1}\log^{(\gamma-1)/(\gamma+1)}(1/\varepsilon).\nonumber
\end{align}
Since the fading is Weibull superexponential we have
\begin{align}
-n\log\PP\left(Z_1>\frac{R^\alpha x}{n\varepsilon}\right)
&\sim\frac{c\kappa}{\varepsilon^{\gamma/(\gamma+1)}\log^{1/(\gamma+1)}(1/\varepsilon)}
\left(\frac{R^\alpha
x\log^{1/(\gamma+1)}(1/\varepsilon)}{\kappa\varepsilon^{1/(\gamma+1)}}\right)^\gamma
\nonumber\\
&=\frac{c(R^\alpha
x)^\gamma}{\kappa^{\gamma-1}}\frac{\log^{(\gamma-1)/(\gamma+1)}(1/\varepsilon)}{\varepsilon^{2\gamma/(\gamma+1)}}.\label{eq:weib2}
\end{align}
Combining \eqref{eq:weib0}, \eqref{eq:weib1} and \eqref{eq:weib2}
we have
\begin{equation}\label{eq:lbW}
\liminf_{\varepsilon\to
0}\frac{\varepsilon^{2\gamma/(\gamma+1)}}{\log^{(\gamma-1)/(\gamma+1)}(1/\varepsilon)}\log\PP(\varepsilon
I_\Lambda> x)\geq-\frac{\gamma\kappa^2}{2(\gamma+1)}
-\frac{c(R^\alpha x)^\gamma}{\kappa^{\gamma-1}}.
\end{equation}
The maximum value of the lower bound is attained at
\[
\kappa=\left(\frac{c(\gamma^2-1)(R^\alpha
x)^\gamma}{\gamma}\right)^{1/(\gamma+1)}.
\]
The claim follows by a straightforward computation substituting
this value of $\kappa$ in \eqref{eq:lbW}.
\\
\noindent$\square$

We conclude this section stating the following immediate corollary
of Theorem \ref{thm:ldW}.
\begin{Corollary}\label{cor:inttruncW}
Under the assumptions of Theorem \ref{thm:ldW},
\begin{equation}\label{resweisup}
\lim_{x\to\infty}\frac{\log\PP(I_\Lambda\geq
x)}{x^{2\gamma/(\gamma+1)}\log^{(\gamma-1)/(\gamma+1)}
x}=-\frac{1}{2}R^{2\alpha\gamma/(\gamma+1)}\left(\frac{\gamma}{\gamma-1}\right)^{(\gamma-1)/(\gamma+1)}(c(\gamma+1))^{2/(\gamma+1)}.
\end{equation}
\end{Corollary}
In this case huge values of the interference are typically obtained as the sum
of a large number of interfering nodes with large signals. This interpretation follows from the proof of Theorem
\ref{thm:ldW}, which establishes that the event $A_\varepsilon^{(n)}$ defined by \eqref{eq:A} with $n$ defined as in \eqref{eq:n}
is a dominating event, as $\varepsilon\to 0$.

Again, we can compare \eqref{resweisup} against its analogue for Poisson networks derived
in \cite{ganesh} and here repeated (see also Proposition 5.2 in \cite{leonardi}):
\begin{Proposition}
If $\mathbf N$ is a homogeneous Poisson process and the fading random variables
are Weibull superexponential as in Theorem \ref{thm:ldW},
\begin{equation}\label{eq:rewe}
\lim_{x\to\infty}\frac{\log\PP(I_\Lambda\geq
x)}{x\log^{(\gamma-1)/\gamma}
x}=-\gamma(\gamma-1)^{-(\gamma-1)/\gamma}c^{1/\gamma}R^\alpha.
\end{equation}
\end{Proposition}
We conclude that also when the fading is Weibull superexponential the tail of the
interference can be significantly reduced by carefully placing transmitting nodes
as regularly as possible. Note that the differences between the terms in
\eqref{resweisup} and \eqref{eq:rewe} vanish as $\gamma \to 1$.
This is hinting at the fact that for exponential or subexponential fading random variables, on the log-scale,
the asymptotic behavior of the tail of the interference becomes insensitive
to the node placement process. This issue will be investigated in Section~\ref{sec:ldpexp}.

\section{Large deviations of the interference: exponential and subexponential fading}\label{sec:ldpexp}

\subsection{Exponential fading}\label{par:exp}

The standing assumptions of this subsection are: \eqref{eq:superexpN}
and the fading random variables $Z_i$, $i\geq 1$, are exponential in the sense
that $-\log\PP(Z_1>z)\sim cz$, for some constant $c>0$.

\begin{Theorem}\label{thm:ldexpsotto}
Under the foregoing assumptions, the family of random variables
$\{\varepsilon I_\Lambda\}_{\varepsilon>0}$ obeys an LDP on
$[0,\infty)$ with speed $\frac{1}{\varepsilon}$ and good rate
function $I_3(x)=c R^{\alpha}x$.
\end{Theorem}

The proof of this theorem is based on the following lemmas whose
proofs are given below.
\begin{Lemma}\label{le:upperexp}
Under the foregoing assumptions, for any $x\geq 0$,
\[
\limsup_{\varepsilon\to 0}\varepsilon\log\PP(\varepsilon
I_\Lambda\geq x)\leq -I_3(x).
\]
\end{Lemma}
\begin{Lemma}\label{le:lowerexp}
Under the foregoing assumptions, for any $x\geq 0$,
\[
\liminf_{\varepsilon\to 0}\varepsilon\log\PP(\varepsilon
I_\Lambda>x)\geq -I_3(x).
\]
\end{Lemma}

\noindent$\it{Proof\,\,of\,\,Theorem\,\,\ref{thm:ldexpsotto}}$ The
claim follows by Proposition \ref{pro:crit} and Lemmas
\ref{le:upperexp} and \ref{le:lowerexp}.
\\
\noindent$\square$
\\

\noindent$\it{Proof\,\,of\,\,Lemma\,\,\ref{le:upperexp}}$ Since
the claim is true if $x=0$, we take $x>0$.
By the assumption on the tail of $Z_1$, one may easily realize
that $\EE[\mathrm{e}^{\delta Z_1}]<\infty$, for any $\delta<c$.
We note here that the inequality \eqref{eq:rev10bis}
holds indeed for general positive random variables $Z_i$, $i\geq 1$, (not necessarily Weibull distributed),
a general point process $\{X_i\}_{i\geq 1}$ (not necessarily a reduced Palm version at the origin of a $\beta$-Ginibre
process), any $\varepsilon,\theta>0$ and any bounded set $\Lambda'$ such that $\Lambda'\supseteq\Lambda$.
Setting $\Lambda'=\Lambda$ and $\theta=(c-\delta)R^\alpha/\varepsilon$ in \eqref{eq:rev10bis}, we deduce
\begin{align}
\PP\left(\varepsilon I_\Lambda\geq
x\right)
\leq\exp\left(-(c-\delta)R^{\alpha}x/\varepsilon+
\log\EE\left[\EE\left[\mathrm{e}^{(c-\delta)Z_1}\right]^{N(\Lambda)}\right]
\right).\nonumber
\end{align}
Therefore by assumption \eqref{eq:superexpN} we have
\begin{equation*}
\limsup_{\varepsilon\to 0}\varepsilon\log\PP\left(\varepsilon I_\Lambda\geq
x\right)\leq-(c-\delta)R^\alpha x.
\end{equation*}
The claim follows letting $\delta$ tend to zero.
\\
\noindent$\square$
\\

\noindent$\it{Proof\,\,of\,\,Lemma\,\,\ref{le:lowerexp}}$ Since
the claim is true if $x=0$, we take $x>0$. Since
$y\in\Lambda^\circ$, there exists $r\in (0,\min\{1,R\})$ such that
$b(y,r)^\circ\subset\Lambda$. For all $\varepsilon>0$, we have
\begin{align}
\PP(\varepsilon I_\Lambda> x)\geq \PP(\varepsilon
I_{b(y,r)^\circ}>x)&
=\PP\left(\sum_{i\geq 1}Z_i\ind_{b(y,r)^\circ}(X_i)>\frac{R^\alpha x}{\varepsilon}\right)\nonumber\\
&\geq\PP\left(Z_1>\frac{R^\alpha
x}{\varepsilon},N(b(y,r)^\circ)\geq 1\right)\nonumber\\
&=\PP\left(Z_1>\frac{R^\alpha
x}{\varepsilon}\right)\PP(N(b(y,r)^\circ)\geq
1),\label{eq:ineqheavy}
\end{align}
where the latter equality follows by the independence of  $N(b(y,r)^\circ)$ and $\{ Z_i\}_{i\ge 1}$.
  
The claim follows by the exponential decay of the tail of $Z_1$,
taking first the logarithm on the above inequality, multiplying
then by $\varepsilon$ and finally letting $\varepsilon$ tend to
zero.
\\
\noindent$\square$

We conclude this section stating the following immediate corollary
of Theorem \ref{thm:ldexpsotto}.
\begin{Corollary}\label{cor:intexp}
Under the assumptions of Theorem \ref{thm:ldexpsotto},
\begin{equation}
\lim_{x\to\infty}\frac{\log\PP(I_\Lambda\geq x)}{x}=-c R^{\alpha}.
\label{fad-exp}
\end{equation}
\end{Corollary}
The fact that under the exponential fading the tail of the interference is given by \eqref{fad-exp},
for any point process satisfying condition \eqref{eq:superexpN}, can be explained by observing that
large values of the interference are typically originated by a single strong interfering contribution
(by \eqref{eq:ineqheavy} clearly emerges that $\{Z_1>R^\alpha
x/\varepsilon\}$ is the dominating event, as $\varepsilon\to 0$.)
In view of these premises, it is reasonable to expect
a similar result also when the distribution of the fading is heavier than
the exponential law. This issue is analyzed for a family of subexponential fading random variables in
the Subsection \ref{par:subexp}.

\subsection{Subexponential fading}\label{par:subexp}

The standing assumptions of this subsection are: \eqref{eq:lightN} and the fading random
variables $Z_i$, $i\geq 1$, are subexponential and such that
\begin{equation}\label{eq:subexp1}
\text{For any $\sigma>0$,
$\lim_{z\to\infty}\frac{\log\overline{F}(\sigma
z)}{\log\overline{F}(z)}=\sigma^\gamma$, for some constant
$\gamma\geq 0$.}
\end{equation}
In particular, note that
the above condition is satisfied if $Z_1$ is subexponential and such
that $-\log\overline{F}(z)\sim cz^\gamma$ (Weibull subexponential
fading) or $-\log\overline{F}(z)\sim c\log z$ (Pareto fading), for
some constants $c>0$ and $\gamma\in (0,1)$.

\begin{Theorem}\label{thm:ldWsubexpsotto}
Under the foregoing assumptions, the family of random variables
$\{\varepsilon I_\Lambda\}_{\varepsilon>0}$ obeys an LDP on
$[0,\infty)$ with speed
$-\log\overline{F}\left(\frac{1}{\varepsilon}\right)$ and rate
function $I_4(0)=0$ and $I_4(x)=R^{\alpha\gamma}x^\gamma$, $x>0$.
\end{Theorem}

The proof of this theorem is based on the following lemmas whose
proofs are given below.
\begin{Lemma}\label{le:upperWsubexp}
Under the foregoing assumptions, for any $x\geq 0$,
\[
\limsup_{\varepsilon\to 0}-\frac{1}{\log\overline{F}\left(\frac{1}{\varepsilon}\right)}\log\PP(\varepsilon
I_\Lambda\geq x)\leq -I_4(x).
\]
\end{Lemma}
\begin{Lemma}\label{le:lowerWsubexp}
Under the foregoing assumptions, for any $x\geq 0$,
\[
\liminf_{\varepsilon\to 0}-\frac{1}{\log\overline{F}\left(\frac{1}{\varepsilon}\right)}\log\PP(\varepsilon
I_\Lambda>x)\geq -I_4(x).
\]
\end{Lemma}

\noindent$\it{Proof\,\,of\,\,Theorem\,\,\ref{thm:ldWsubexpsotto}}$
The claim follows by Proposition \ref{pro:crit} and Lemmas
\ref{le:upperWsubexp} and \ref{le:lowerWsubexp}.
\\
\noindent$\square$
\\

\noindent$\it{Proof\,\,of\,\,Lemma\,\,\ref{le:upperWsubexp}}$
Since the claim is true if $x=0$, we take $x>0$. By assumption $N(\Lambda)$
has a convergent Laplace transform in a right neighborhood of
zero, therefore since $Z_1$ is subexponential by e.g. Lemma 2.2 p.
259 in \cite{asmussen} it follows
\begin{equation}\label{eq:subexp}
\PP\left(\sum_{i=1}^{N(\Lambda)}Z_i\geq
x\right)\sim\EE[N(\Lambda)]\overline{F}(x),\quad\text{as
$x\to\infty$.}
\end{equation}
We note here that the inequality \eqref{eq:rev10ter} holds indeed for general positive random variables $Z_i$, $i\geq 1$, (not necessarily with bounded support),
a general point process $\{X_i\}_{i\geq 1}$ (not necessarily a reduced Palm version at the origin of a $\beta$-Ginibre
process) and any $\varepsilon,x>0$. By \eqref{eq:rev10ter} and \eqref{eq:subexp} easily follows that
\begin{align}
&\limsup_{\varepsilon\to 0}-\frac{1}{\log\overline{F}(1/\varepsilon)}\log\PP(\varepsilon
I_\Lambda\geq x)\nonumber\\
&\leq\limsup_{\varepsilon\to 0}
-\frac{1}{\log\overline{F}(1/\varepsilon)}
\log\PP\left(\sum_{i\geq 1}Z_i\ind_{\Lambda}(X_i)\geq\frac{
R^\alpha
x}{\varepsilon}\right)\nonumber\\
&=\limsup_{\varepsilon\to 0}
-\frac{1}{\log\overline{F}(1/\varepsilon)}
\log\left(\EE[N(\Lambda)]\overline{F}\left(\frac{R^\alpha
x}{\varepsilon}\right)\right)=-R^{\alpha\gamma}
x^\gamma,\nonumber
\end{align}
where the latter equality is consequence of condition
\eqref{eq:subexp1}.
\\
\noindent$\square$
\\

\noindent$\it{Proof\,\,of\,\,Lemma\,\,\ref{le:lowerWsubexp}}$
Since the claim is true if $x=0$, we take $x>0$. Arguing as in the
proof of Lemma \ref{le:lowerexp} we have the inequality
\eqref{eq:ineqheavy}. The claim follows by the subexponential
decay of the tail of $Z_1$, taking first the logarithm on the
inequality \eqref{eq:ineqheavy}, multiplying then by
$-\frac{1}{\log\overline{F}(1/\varepsilon)}$ and finally letting
$\varepsilon$ tend to zero.
\\
\noindent$\square$

We conclude this section stating the following immediate corollary
of Theorem \ref{thm:ldWsubexpsotto}.
\begin{Corollary}\label{cor:intWsubexp}
Under the assumptions of Theorem \ref{thm:ldWsubexpsotto},
\[
\lim_{x\to\infty}\frac{\log\PP(I_\Lambda\geq x)}{\log\overline{F}(x)}=
R^{\alpha\gamma}.
\]
\end{Corollary}
Note that also when the fading is subexponential
large values of the interference are due to a single strong interfering node,
for any point process which satisfies \eqref{eq:lightN}.
\section{Conclusions} \label{sec:conc}
The results of this paper contribute to better understand the reliability of large scale wireless networks.
We proved asymptotic estimates, on the log-scale, for the tail of the interference in a network whose nodes
are placed according to a $\beta$-Ginibre process (wiith $0<\beta \le 1$)  and the fading random variables
are bounded or Weibull superexponential. We gave also asymptotic estimates, on the log-scale, for the tail of the interference in a network whose nodes
are placed according to a general point process and the fading random variables
are exponential or subexponential.
The  results, summarized in Tables ~\ref{Table:summ} and ~\ref{Table:summ1},
 show the emergence of two different regimes (for the ease of comparison results for the Poisson model
under bounded or Weibull superexponential fading are reported in Table~\ref{Table:summPoisson}).
When the fading variables are bounded  or Weibull superexponential,
the tail of the interference heavily depends on the
the node spatial process. Instead, when the fading variables are exponential or subexponential, the tail of the interference
is essentially insensitive to
the distribution of nodes, as long as the number of nodes is guaranteed to be light-tailed.

\begin{table}
\begin{center}
\begin{tabular}{|l|l|l|}
\hline
Fading distribution & Speed & Rate function \\
\hline
Bounded & $\frac{1}{\varepsilon^2}\log\left(\frac{1}{\varepsilon}\right)$ & $\frac{R^{2\alpha}x^2}{2B}$ \\
\hline
Weibull superexponential & $\varepsilon^{-\frac{2\gamma}{\gamma+1}}\log^{\frac{\gamma-1}{\gamma+1}}\left(\frac{1}{\varepsilon}\right)$ &
$\frac{1}{2}R^{\frac{2\alpha\gamma}{\gamma+1}}\left(\frac{\gamma}{\gamma-1}\right)^{\frac{\gamma-1}{\gamma+1}}(c(\gamma+1))^{\frac{2}{\gamma+1}}x^{\frac{2\gamma}{\gamma+1}}$ \\
\hline
\end{tabular}
\caption{LDPs of the family $\{\varepsilon I_\Lambda\}$ \label{Table:summ}, when the nodes are
distributed according to a $\beta$-Ginibre process, $0<\beta\leq 1$.}
\end{center}
\end{table}

\begin{table}
\begin{center}
\begin{tabular}{|l|l|l|}
\hline
Fading distribution & Speed & Rate function \\
\hline
Bounded & $\frac{1}{\varepsilon}\log\left(\frac{1}{\varepsilon}\right)$ & $\frac{R^{\alpha}x}{B}$ \\
\hline
Weibull superexponential & $1/\varepsilon\log^{1-\frac{1}{\gamma}}\left(\frac{1}{\varepsilon}\right)$ &
$\gamma(\gamma-1)^{\frac{1}{\gamma}-1}c^{\frac{1}{\gamma}}R^{\alpha} x $ \\
\hline
\end{tabular}
\caption{LDPs of the family $\{\varepsilon I_\Lambda\}$ \label{Table:summPoisson}, when the nodes are
distributed according to a Poisson process.}
\end{center}
\end{table}

\begin{table}
\begin{center}
\begin{tabular}{|l|l|l|}
\hline
Fading distribution & Speed & Rate function \\
\hline
Exponential & $1/\varepsilon$ & $cR^\alpha x$ \\
\hline
$\log\overline{F}(\sigma x)\sim\sigma^\gamma\log\overline{F}(x)$ & $-\log\overline{F}(1/\varepsilon)$ & 0 if $x=0;\;$ $R^{\alpha\gamma} x^\gamma$ if $x>0$ \\
\hline
\end{tabular}
\caption{LDPs of the family $\{\varepsilon I_\Lambda\}$ \label{Table:summ1}, when the number of nodes
is light-tailed. Here $\overline{F}=1-F$, being $F$ the distribution function of the fading and $\sigma>0$, $\gamma\geq 0$.}
\end{center}
\end{table}

\section*{Appendix}

\subsection*{Proof of Proposition \ref{pro:crit}}

Let $F$ be a closed subset of $[0,\infty)$ and let $x$ denote the
infimum of $F$. Since $I$ is increasing, $I(x) =
\inf_{y\in F} I(y)$. Since $F$ is contained in $[x,\infty)$, by the large deviation
upper bound for closed half-intervals $[x,\infty)$ we deduce
\begin{eqnarray*}
\limsup_{\varepsilon\to 0}\frac{1}{v(\varepsilon)}\log\PP(V_{\varepsilon}
\in F) &\le& \limsup_{\varepsilon \to 0} \frac{1}{v(\varepsilon)}\log
\PP(V_{\varepsilon}\ge x)\\
&\le& -I(x)=-\inf_{y\in F} I(y).
\end{eqnarray*}
This establishes the large deviation upper bound for arbitrary
closed sets.

Now, let $G$ be an open subset of $[0,\infty)$.
Suppose first that $0\notin G$. Since $\inf_{y\in G}
I(y)<\infty$, for arbitrary $\delta>0$, we
can find $x\in G$ such that $I(x)\le\inf_{y\in G}
I(y)+\delta$. Since $G$ is
open, we can also find $\eta>0$ such that $(x-\eta,x+\eta)
\subset G$. By the large deviation bounds on half-intervals we have
\begin{equation*}
\liminf_{\varepsilon\to 0}\frac{1}{v(\varepsilon)}\log P(V_{\varepsilon}>
x-\eta) \ge -I(x-\eta)
\end{equation*}
and
\begin{equation*}
\limsup_{\varepsilon\to 0} \frac{1}{v(\varepsilon)}\log\PP(V_{\varepsilon}
\ge x+\eta) \le -I(x+\eta),
\end{equation*}
and by the monotonicity of $I$ we deduce $I(x-\eta)\le I(x+\eta)$. Consequently, after an easy computation we get
$$
\liminf_{\varepsilon \to 0} \frac{1}{v(\varepsilon)}\log(\PP(V_{\varepsilon}>x-\eta)-\PP(V_{\varepsilon} \ge x+\eta))
\ge -I(x-\eta).
$$
Note that
\begin{equation*}
\PP(V_{\varepsilon} \in G) \ge \PP(V_{\varepsilon}\in (x-\eta,x+\eta)) =
\PP(V_{\varepsilon}>x-\eta)-\PP(V_{\varepsilon} \ge x+\eta),
\end{equation*}
and so
$$
\liminf_{\varepsilon \to 0} \frac{1}{v(\varepsilon)}\log\PP(V_{\varepsilon}
\in G) \ge -I(x-\eta).
$$
Since $I$ is continuous on $(0,\infty)$, by letting $\eta$ tend to zero we get
$$
\liminf_{\varepsilon \to 0} \frac{1}{v(\varepsilon)}\log\PP(V_{\varepsilon}
\in G) \ge -I(x) \ge -\inf_{y\in G} I(y) - \delta,
$$
where the latter inequality follows by the choice of $x$. The
large deviation lower bound for arbitrary open sets not containing the origin
follows letting $\delta$ tend to zero. If $0\in G$, then, since $G$ is open, there is an $\eta>0$ such
that $[0,\eta) \subset G$. Hence,
$$
\PP(V_{\varepsilon} \in G) \ge 1-\PP(V_{\varepsilon} \ge \eta).
$$
By similar arguments to the above, we can show that
$$
\liminf_{\varepsilon \to 0} \frac{1}{v(\varepsilon)}\log\PP(V_{\varepsilon}
\in G) \ge 0.
$$
Since $I$ is
increasing we have $\inf_{y\in G} I(y) = I(0) = 0$, and the proof is completed.

\subsection*{Proof of Lemma \ref{le:ginibre}}
\noindent{\it Proof\,\,of\,\,$(i)$} By Theorem 6 in \cite{krishnapur}, for any fixed $r>0$ and
$x_0\in\C$, we have
\begin{equation}\label{eq:krishnapur}
\PP\left(\sum_{i\geq 1}\ind_{b(x_0,r)}(V_i)\geq
m\right)=\mathrm{e}^{-\frac{1}{2}m^2\log m(1+o(1))}
\end{equation}
(note that the processes $\{X_i\}_{i\geq 1}$ and $\{V_i\}_{i\geq 1}$ are different
and so a priori one can not say that the tails of $\PP\left(\sum_{i\geq 1}\ind_{b(x_0,r)}(V_i)\geq
m\right)$ and $\PP\left(N(b(x_0,r))\geq
m\right)$ are equal.)
Since the Ginibre process is stationary, so is the independently thinned process and
thus it suffices to check \eqref{eq:krishnapurtinnato} with $x_0=O$.
By \eqref{eq:krishnapur} we have
\[
\PP\left(\sum_{i\geq 1}\ind_{b(O,r)}(V_i)\ind_{A_i}\geq
m\right)\leq\mathrm{e}^{-\frac{1}{2}m^2\log m(1+o(1))}.
\]
It remains to check the matching lower bound. The function
\[
r\mapsto\PP\left(\sum_{i\geq 1}\ind_{b(O,r)}(V_i)\ind_{A_i}\geq
m\right)
\]
is clearly nondecreasing. Since we are going to check the lower bound,
we may assume $0<r<1$. We have
\begin{align}
\PP\left(\sum_{i\geq 1}\ind_{b(O,r)}(V_i)\ind_{A_i}\geq
m\right)&\geq\PP\left(\ind_{b(O,r)}(V_i)\ind_{A_i}=1,\,\,
\text{$\forall$ $i=1,\ldots,m$}\right)\nonumber\\
&=\PP\left(|V_i|<r,\,A_i,\,\,\text{$\forall$
$i=1,\ldots,m$}\right)\nonumber\\
&=\PP(A_1)^{m}\PP(|V_i|<r,\,\,
\text{$\forall$ $i=1,\ldots,m$}).\label{eq:rev7}
\end{align}
By Theorem 1.1 in \cite{kostlan} (see also Theorem 4.7.3 p. 73 in \cite{hough}) the set
$\{|V_i|\}_{i\geq 1}$ has the same distribution as the set
$\{\rho_i\}_{i\geq 1}$, where the random variables $\rho$ are
independent and $\rho_i^2$ has the Gamma($i$,1) distribution for
every $i\geq 1$. Hence $\rho_i^2$ has the same distribution of
$\xi_{i1}+\ldots+\xi_{ii}$, where the random variables $\{\xi_{jk}\}_{j,k\geq 1}$ are
independent and have the Exponential($1$) distribution. So
\begin{align}
\PP\left(|V_i|<r,\,\,\text{$\forall$
$i=1,\ldots,m$}\right)&=\PP\left(\rho_i^2<r^2,\,\,
\text{$\forall$ $i=1,\ldots,m$}\right)\nonumber\\
&=\PP\left(\sum_{k=1}^{i}\xi_{ik}<r^2,\,\,
\text{$\forall$ $i=1,\ldots,m$}\right)\nonumber\\
&=\prod_{i=1}^{m}\PP\left(\sum_{k=1}^{i}\xi_{ik}<r^2\right)\label{eq:rev4}\\
&\geq\prod_{i=1}^{m}\PP\left(\xi_{ik}<r^2/i,\,\,
\text{$\forall$ $k=1,\ldots,i$}\right)\nonumber\\
&=\prod_{i=1}^{m}\prod_{k=1}^{i}\PP\left(\xi_{ik}<\frac{r^2}{i}\right)
=\prod_{i=1}^{m}
\left(1-\mathrm{e}^{-\frac{r^2}{i}}\right)^{i}\label{eq:rev5}\\
&\geq\prod_{i=1}^{m}\left(\frac{r^2}{2i}\right)^{i},\label{eq:rev6}
\end{align}
where \eqref{eq:rev4} and the first equality in \eqref{eq:rev5}
follow by the independence of the random variables $\xi_{jk}$
and the inequality \eqref{eq:rev6} is a consequence of the fact that $0<r^2/i<1$ for any $i=1,\ldots,m$
and $1-\mathrm{e}^{-x}\geq x/2$ for $0<x<1$. Combining \eqref{eq:rev7} and \eqref{eq:rev6} and using the elementary inequality
\[
\PP(A_1)^m\geq\PP(A_1)^{m(m+1)/2}=\prod_{i=1}^{m}\PP(A_1)^i
\]
we have
\begin{equation}\label{eq:palm2app}
\PP\left(\sum_{i\geq 1}\ind_{b(O,r)}(V_i)\ind_{A_i}\geq
m\right)\geq
\prod_{i=1}^{m}\left(\frac{\PP(A_1)r^2}{2i}\right)^{i}.
\end{equation}
A straightforward computation shows that
\begin{align}
\prod_{i=1}^{m}
\left(\frac{\PP(A_1)r^2}{2i}\right)^{i}&=\left(\frac{\PP(A_1)r^2}{2}\right)^{\frac{m(m+1)}{2}}
\exp\left(-\sum_{i=1}^{m}i\log i\right)\nonumber\\
&\geq\left(\frac{\PP(A_1)r^2}{2}\right)^{\frac{m(m+1)}{2}}
\exp\Biggl(-\frac{1}{2}(m+1)^2\log(m+1)+\frac{(m+1)^2}{4}-\frac{1}{4}\Biggr)\label{eq:lbbd3app}\\
&=\mathrm{e}^{-\frac{1}{2}m^2\log m(1+o(1))},\nonumber
\end{align}
where the inequality in \eqref{eq:lbbd3app} follows by the elementary relation:
\[
\sum_{i=1}^{m}i\log
i\leq\frac{1}{2}(m+1)^2\log(m+1)-\frac{(m+1)^2}{4}+\frac{1}{4},\qquad\text{$m\geq
1$.}
\]
The proof is completed.
\\
\noindent{\it Proof\,\,of\,\,$(ii)$}
Letting $\{U\}\cup\{U_i\}_{i\geq 1}$ denote a sequence of independent
random variables uniformly distributed on $[0,1]$ and
$Z$ denote a random variable distributed as $Z_1$, and assuming that the random variables
$\{U,Z\}\cup\{U_i\}_{i\geq 1}$ are independent of all the other
random quantities.
For any bounded and measurable set $\Lambda'\subset\mathbb C$, by Lemma \ref{le:idlaw} we have
\begin{align}
N(\Lambda')+\ind_{\Lambda'}(\sqrt\beta G)\ind\{U<\beta\}
&\overset{law}=\sum_{i\geq 1}\ind_{\Lambda'}(\sqrt\beta V_i)\ind\{U_i<\beta\}\nonumber\\
=\sum_{i\geq 1}\ind_{\Lambda'/\sqrt\beta}(V_i)\ind\{U_i<\beta\},\label{eq:goldman1}
\end{align}
where the symbol $\overset{law}=$ denotes the identity in law.
Note that
\begin{align}
&\PP\left(N(b(x_0,r))+\ind_{b(x_0,r)}(\sqrt\beta G)\ind\{U<\beta\}\geq m+1\right)\nonumber\\
&\,\,\,\,\,\,\,\,\,\,\,\,\,\,\,\leq\PP\left(N(b(x_0,r))+\ind_{b(x_0,r)}(\sqrt\beta G)\ind\{U<\beta\}\geq m+\ind_{b(x_0,r)}(\sqrt\beta G)\ind\{U<\beta\}\right)\nonumber\\
&\,\,\,\,\,\,\,\,\,\,\,\,\,\,\,=\PP\left(N(b(x_0,r))\geq
m\right).\nonumber
\end{align}
Combining \eqref{eq:goldman1} (with $\Lambda'=b(x_0,r)$) and this latter relation, we have
\begin{align}
&\PP\left(\sum_{i\geq 1}\ind_{b(x_0/\sqrt\beta,r/\sqrt\beta)}(V_i)\ind\{U_i<\beta\}\geq
m+1\right)\nonumber\\
&\leq\PP(N(b(x_0,r))\geq m)\nonumber\\
&\leq\PP(N(b(x_0,r))+\ind_{b(x_0,r)}(\sqrt\beta G)\ind\{U<\beta\}\geq m)\nonumber\\
&\leq\PP\left(\sum_{i\geq 1}\ind_{b(x_0/\sqrt\beta,r/\sqrt\beta)}(V_i)\ind\{U_i<\beta\}\geq m\right).
\nonumber
\end{align}
The claim follows by
\eqref{eq:krishnapurtinnato}.

\subsection*{Proof of Lemma \ref{le:laplacemean}}
By \eqref{eq:goldman1} we have
\begin{align}
\EE[N(\Lambda')]&\leq\EE\left[N(\Lambda')+\ind_{\Lambda'}(\sqrt\beta G)\ind\{U<\beta\}\right]\nonumber\\
&=\EE\left[\sum_{i\geq 1}\ind_{\Lambda'/\sqrt\beta}(V_i)\ind\{U_i<\beta\}\right]\nonumber\\
&\leq\EE\left[\sum_{i\geq 1}\ind_{\Lambda'/\sqrt\beta}(V_i)\right]\nonumber\\
&=\sum_{n\geq
1}\kappa_n(\Lambda'/\sqrt\beta)
<\infty,
\label{eq:shirai}
\end{align}
where \eqref{eq:shirai} follows by e.g.
Proposition 2.3 in \cite{shirai} and formula (3.41) in
\cite{takahashi}. Now we prove \eqref{eq:laplace}. Let $\theta\geq
0$ be arbitrarily fixed. We start checking that
\[
\prod_{n\geq
1}(1+(\mathrm{e}^{\theta}-1)\kappa_n(\Lambda'/\sqrt\beta))<\infty.
\]
We have
\begin{align}
\log\prod_{n\geq 1}(1+(\mathrm{e}^{\theta}-1)\kappa_n(\Lambda'/\sqrt\beta))
&=\sum_{n\geq 1}\log(1+(\mathrm{e}^{\theta}-1)\kappa_n(\Lambda'/\sqrt\beta))\nonumber\\
&\leq(\mathrm{e}^{\theta}-1)\sum_{n\geq
1}\kappa_n(\Lambda'/\sqrt\beta)<\infty,\label{eq:log1}
\end{align}
where in \eqref{eq:log1} we used the inequality $x\geq\log(1+x)$,
$x\geq 0$, and \eqref{eq:trace}. Finally, we prove the first inequality in
\eqref{eq:laplace}. Using again \eqref{eq:goldman1}, for any
$\theta\geq 0$, we have
\begin{align}
\EE[\mathrm{e}^{\theta N(\Lambda')}]\leq\EE\left[\mathrm{e}^{\theta
(N(\Lambda')+\ind_{\Lambda'}(\sqrt\beta G)\ind\{U<\beta\})}\right]
&\leq\EE\left[\exp\left(\theta\sum_{i\geq 1}\ind_{\Lambda'/\sqrt\beta}(V_i)\right)\right]\nonumber\\
&=\prod_{n\geq 1}(1+(\mathrm{e}^{\theta}-1)\kappa_n(\Lambda'/\sqrt\beta)),\nonumber
\end{align}
where the latter equality follows by e.g. Proposition 2.2 in
\cite{shirai}. The proof is completed.

\end{document}